\documentclass{svmult}
\usepackage{amssymb,lh}
\newcommand{\Q}{{\mathbb Q}}

\begin{document}

\title*{Conformal Field Theory and Torsion Elements of the Bloch Group}

\author{Werner Nahm}
\institute{Dublin Institute for Advanced Studies
\texttt{wnahm@stp.dias.ie}}

\maketitle

\begin{abstract}
We argue that rational conformally invariant quantum field theories in
two dimensions are closely related to torsion elements of the algebraic
K-theory group $K_3(\C)$. If such a theory has an integrable matrix
perturbation with purely elastic scattering matrix, then the partition 
function has a canonical sum representation. Its asymptotic behaviour is
given in terms of the solution of an algebraic equation which can be read
off from the scattering matrix. The solutions yield torsion elements of an
extension of the Bloch group which seems to be equal to $K_3(\C)$. These
algebraic equations are solved for integrable models given by arbitrary
pairs of equations are solved for integrable models given by arbitrary
pairs of A-type Cartan matrices. The paper should be readable by 
mathematicians.

\end{abstract}

\section{Introduction}

We will study a relation between certain integrable quantum field theories in
two dimensions and the algebraic $K$-theory of the complex numbers. As was
acknowledged in the first publication \cite{NRT93}, discussions with
A.~Goncharov, A.N.~Kirillov and D.~Zagier were
essential, and the results appear to be well worth of the common attention
of physicists and mathematicians. Some developments in physics and to a lesser
degree in mathematics were stimulated, but progress was slow and a fresh start
by young researchers in both fields may be in order.

For a long time few mathematicians attempted to cope with the obscurities of
quantum field theory, but this has started to change. In particular, the much
studied vertex operator algebras capture many features of the conformally 
invariant quantum field theories in two dimensions. The present investigation
will take us a small step farther, since we have to consider integrable 
massive perturbations of these conformally invariant field theories (CFTs). 
On the other hand, only rational CFTs will be investigated. For theories with  
continuous parameters, only special points in their moduli space can be
rational, but those points may allow a more complete understanding than 
generic ones and can serve to explore a neighbourhood in CFT moduli space by 
perturbation theory.

Physicists have a tradition of openness to all kind of mathematics, but
algebraic $K$-theory has not been very popular, in part because its definition 
is very abstract. Much of it is captured by the Bloch group, however, which
is easy to work with. The Bloch group 
$B(K)$ of a field $K$ is an abelian group given by the group cohomology of
$SL(2,K)$. It is a subquotient of the group $\Z[K^\times]$ (the free abelian 
group which has one generator $[x]$ for each non-zero $x\in K$). Thus the 
elements of $B(K)$ can be labelled by elements of $\Z[K^\times]$, though not
in a unique way. The torsion sugroup of an abelian group is the subgroup
consisting of the elements of finite order. The torsion subgroup of 
$\Z[K^\times]$ is trivial, but in the Bloch groups relations like 
$[1/2]+[1/2]=0$ may introduce non-trivial torsion. We shall study a map from 
finite order elements to the central charges and scaling dimensions of 
conformal field theories. The mapping is provided by the dilogarithm, which is
a function known for unexpected appearances in diverse fields in mathematics 
and physics, notably perturbative quantum field theory (see Weinzierl's talk 
\cite{W03}), the classification of hyperbolic three-manifolds, and
algebraic $K$-theory. In perturbation theory and algebraic $K$-theory, higher
polylogarithms appear, too, but it is not known if there is a connection.

Integrable quantum field theories have been treated by the thermodynamic
Bethe ansatz, and in this context the dilogarithm formulas were 
discovered \cite{Z91}, at first for real argument. For extensions to complex
arguments, see \cite{M91, N93, DT98}. 
The original derivation of the formulas was cumbersome, but we shall study a 
much easier one. In particular, we do not need any thermodynamics. The argument
is presented in the second half of section three. Specialists may start to read
there, since the preceding part of the article concerns well known facts and 
is quite elementary. 

Indeed, much of this article is aimed at mathematicians who want to see 
quantum field theory in an understandable language and for young physicists 
who do not want to learn it as an exclusive art. Thus the next section is
a pedagogical introduction into the most elementary aspects of quantum field
theory. No history of the ideas is given, all results are standard, and almost 
all calculations should be easily reproducible by the reader. Functional 
integrals are irrelevant. The language of vertex operator algebras will only
be mentioned in this introduction, since I think that it is a bit far from
God's book of optimal proofs and since the standard physics formulation is far
better suited for the discussion of perturbations which break conformal
invariance. Physically, vertex operator algebras describe the operator product
expansions of holomorphic and anti-holomorphic fields, which will of course be
central to the later discussions. Free fermions and the simplest minimal models
will be considered in enough detail to obtain a first concrete understanding
of the relations between some quantum field theories and algebraic $K$-theory.  
Hopefully, the newcomer who works through this material will find it helpful  
for the study of more complex cases. 

A rational CFT has a holomorphic and an anti-holomorphic vertex operator 
algebra. They have finite sets of characters $\chi_i$, $\bar\chi_j$, resp.
The $\chi_i$ are holomorphic functions of the complex upper half plane and  
have the form 
\beq{chi}
\chi_i(\tau)=q^{h_i-c/24}\sum_{n=0}^\infty a_{in} q^n,
\eeq
with $q =\exp(2\pi i\tau)$, rational $c,h_i$ and positive integers $a_{in}$.
For the distinguished vacuum character $\chi_0$ one has $h_0=0$, $a_{00}=1$.
The $q$ exponents $h_i+n$ are the holomorphic conformal dimensions of the CFT,
and $c$ is its holomorphic central charge. The minimal value among the 
differences $c-24 h_i$ is called the effective central charge $c_{eff}$. In  
unitary theories $h_i>0$ for $i\neq 0$, such that $c_{eff}=c$.
After complex conjugation, the properties of the $\bar\chi_j$ are analogous, 
and in many cases these characters are indeed the complex conjugates of the 
$\chi_i$. We are interested in massive perturbations without holomorphic
or anti-holomorphic fields. In CFTs with such perturbations the
holomorphic and anti-holomorphic central charges have to be equal.

The partition function of a bosonic rational CFT has the form
\beq{Z}
Z=\sum_{i,j} n_{ij} \chi_i\bar\chi_j,
\eeq
where $n_{00}=1$ and all $n_{ij}$ are positive integers (possibly zero). 
As a consequence of conformal invariance, the partition function $Z$ is 
invariant under the modular group $SL(2,\Z)$, the elements ${AB\choose CD}$
of which act on $\tau$ in the form $\tau\mapsto (A\tau+B)/(C\tau+D)$.
Consequently, the $\chi_i$ form a vector valued representation of the
modular group, such that they transform into a linear combinations of each
other, with constant coeffients. Functions which generate a finite dimensional
vector space under $SL(2,\Z)$ transformations will be called modular. The
terminology is not quite the standard one, but this shouldn't matter, since
the individual $\chi_i$ are also expected to be invariant under a congruence 
subgroup of $SL(2,\Z)$. We will not need this property, however. Fermionic
CFTs have slightly more structure in their partition functions, but their
characters are modular, too. 

The property which is essential for us is the behaviour under the particular 
modular transformation $\tau\mapsto -1/\tau$, namely the fact that the
$\chi_i$ can be written in the form
\beq{tildechi}
\chi_i(\tau)=\sum_k \tilde a_{ik} \tilde q^k,
\eeq
with $\tilde q =\exp(-2\pi i/\tau)$, where the sum goes over rational numbers 
of the form $k=h_j-c/24+n$, $n\in\N$. For small $\tau$ the dominating exponent 
is $k_0=-c_{eff}/24$, and the corresponding coefficient $\tilde a_{ik_0}$ is 
real and strictly positive for all $i$. All coefficients $\tilde a_{ik}$ turn  
out to be algebraic numbers, but they will not be studied in this article.

Apart from the factor $q^{h_i-c/24}$ the $\chi_i$ also can be described
in a combinatorial way, sometimes in terms of very classical
combinatorics. We shall see in the next sections how this comes about.
For the moment let us just list some of the relevant functions.
The partitions of natural numbers into $n$ distinct summands have the
generating function
$$\frac{q^{(n^2+n)/2}}{(q)_n}\,,$$
where $(q)_n=(1-q)(1-q^2)\cdots (1-q^n)$. Thus

$$\sum_{n\in \N} \frac{q^{(n^2+n)/2}}{(q)_n}=\prod_{n=1}^\infty (1+q^n)$$
(in our dialect, $\N=\{0,1,\ldots\}$).
Analogously, the partitions into distinct odd summands yield
$$\sum_{n\in \N} \frac{q^{n^2}}{(q^2)_n}=\prod_{n=1}^\infty (1+q^{2n+1}).$$

Up to factors $q^{1/24}$ and $q^{-1/24}$, resp., these are modular functions.
Another example of equalities between such sums and products is provided by
the Rogers-Ramanujan identities 
\begin{eqnarray}\la{RR}
\sum_{n\in \N} \frac{q^{n^2+n}}{(q)_n}&=&\prod_{n\equiv 2,3\,\mod \,5}
(1-q^n)^{-1}\nonumber\\
\sum_{n\in \N} \frac{q^{n^2}}{(q)_n}&=&\prod_{n\equiv 1,4\,\mod \,5}
(1-q^n)^{-1}.
\end{eqnarray}
Again these are modular functions, up to factors $q^{11/60}$ and
$q^{-1/60}$, resp. A first generalization is provided by the Andrews-Gordon
identities \cite{A76}.  

A more surprising example concerns the partitions of the natural numbers into
distinct half-integral summands, for which one obtains \cite{KM93}
\beq{e8}
\sum_{n\in \N} \frac{q^{2n^2}}{(q)_{2n}}=
\sum_{m\in\N^8} \frac{q^{mCm}}{(q)_m}.
\eeq
Here $m=(m_1,\ldots,m_8)$, $(q)_m=(q)_{m_1}\cdots (q)_{m_8}$, and
$$C=\left(\begin{array}{lccccccc}2&3&4&5&6&4&3&2\\3&6&8&10&12&8&6&4\\
4&8&12&15&18&12&9&6\\5&10&15&20&24&16&12&8\\
6&12&18&24&30&20&15&10\\4&8&12&16&20&14&10&7\\
3&6&9&12&15&10&8&5\\2&4&6&8&10&7&5&4\end{array}\right)$$
is the inverse of the Cartan matrix of the exceptional Lie algebra
$E_8$. Formula \req{e8} is still unproven, but easy to check to high
order. Much work has been done on the combinatorial side, see e.g. 
\cite{BM98}, \cite{ABD03}. The present article can only give a few hints
in this direction.

The combinatorial structures have deep roots in quantum field theory,
In section 3 we shall obtain the form
\beq{chiQ}
\chi_i(\tau)= \sum_m  \frac{q^{Q_i(m)}}{(q)_m} 
\eeq
with

$$Q_i(m)=mAm/2+b_im+h_i-c/24$$
for charcters of certain CFTs with integrable deformations. Note that the 
matrix $A$ is the same for all characters $\chi_i$ of a given CFT.
The form comes from the Bethe ansatz, which is a well tested conjecture
concerning the solution of integrable theories. The derivation is very
simple and straightforward, but seems to be new. 

In section 4 we consider general sums of the form $\sum_m q^{Q(m)}/(q)_m$,
where $Q(m)=mAm/2+bm+h$ has rational coefficients. We shall see that
such a function can only be modular when all solutions of the system
of equations
\beq{Alog}
\sum_j A_{ij} \log(x_j) = \log(1-x_i)
\eeq
yield elements $\sum_i [x_i]$ of finite order in the Bloch group of the
algebraic numbers, or more precisely in a certain extension of it which takes
into account the multivaluedness of the logarithm. The argument is incomplete,
but it should not be too difficult to make it rigorous.

Note that exponentiation of eq. \req{Alog} yields algebraic equations for  
$x_i$, with a finite number of solutions. In important special cases the $x_i$
are real and positive. In this case the multivaluedness of the logarithm is 
irrelevant and one can work with the Bloch group $B(\bar\Q^+)$ of the field 
$\bar\Q^+$ of all real algebraic numbers or equivalently with 
$B(\R)=B(\bar\Q^+)$.
This group has a torsion subgroup isomorphic to $\Q/\Z$, which naturally
encodes conformal dimensions $h_i$ modulo the integers. It is not really
necessary to consider the fields $\R$ and $\C$ instead of their algebraic
subfields, but for physicists it certainly is natural.

In general it may seem natural to work with the Bloch group $B(\C)=B(\bar\Q)$, 
where $\bar\Q$ is the field of all algebraic numbers, but here
it is inappropriate to forget about the multivaluedness of the logarithm,
since the torsion subgroup of $B(\C)$ is trivial \cite{S86, S89}. Our
quantum field theory context leads to an extension $\hat B(\C)$ of $B(\C)$,
however, in which $\C$ is replaced by a cover $\hat\C$ of $\C-\{0,1\}$
on which $\log(x)$ and $\log(1-x)$ are single-valued. The usual Bloch group 
$B(\C)$ is just the quotient of $\hat B(\C)$ by its torsion subgroup, which
again is isomorphic to $\Q/\Z$. The group $\hat B(\C)$ is a very natural 
object in algebraic K-theory and emerged in the study of hyperbolic 
three-manifolds in the context of Thurston's program \cite{NZ85, GZ00, N03}. 

In Thurston's program, hyperbolic three-manifolds are described by
triangulations into tetrahedra in the standard hyperbolic space $\H^3$.
Every triangulation represents an element of $\hat B(\C)$, and a change of the 
triangulation gives another $\Z[\hat\C]$ representation of the same element.
Moreover, there is a map $D: B(\C)\rightarrow \R$ which yields the volume of 
the manifold when the curvature of $\H^3$ is normalized to $-1$, and this 
volume is an invariant of the three-manifold. Now $D$ is essentially the 
imaginary part of the Rogers dilogarithm $L$, which provides a natural group 
homomorphism $L: \hat B(\C)\rightarrow \C /\Z(2)$, as discussed in Zagier's 
talk. The real part of $L$ yields the Chern-Simons invariant of the manifold.
The symbol $\Z(N)$ stands for $(2\pi i)^N\Z$ and indicates that the right 
context for both the classification of three-manifolds and of our CFTs should 
be the theory of motives.

In the quantum field theoretical situation, we deal with the torsion 
subgroup of $\hat B(\C)$, for which $(2\pi i)^2L$ takes values in $\Q/\Z$.
These values yield the conformal dimensions of the fields of the theory (more
precisely the exponents $h_i-c/24$).
The fact that they are only obtained modulo the integers is natural, since
each $\chi_i$ yields a whole family $h_i+n$ of conformal dimensions.
Nevertheless, $h_i\in\Q$ is the smallest one among them, and a refinement
of the K-theory should yield this number. Similarly, in the study of
three-manifolds it should be possible to remove the ambiguity of the 
Chern-Simons invariant. 

The geometrical ideas just mentioned have been used to derive dilogarithm
identities in conformal field theory \cite{GT96}, but the
construction of a more direct link between Thurston's program and conformal
field theory may be another task for the future. It is tempting to relate the 
conformal field theories in two dimensions to topological ones in three
dimensions and to use the latter for calculating invariants of the
three-manifolds.

Here we will be concerned by more elementary issues, however. In section 5 we 
will take a closer look at eqs. \req{Alog}. One knows many matrices $A$ for 
which these equations yield torsion
elements of the extended Bloch group. The best known examples are related to 
Cartan matrices, or equivalently to Dynkin diagrams. For Dynkin diagrams
of A type, we will find all solutions of the equations, whereas so far
only one solution was known \cite{K87, KR87}, up to the action of the Galois 
group. It turns out that $x_i$ are rational linear combinations of roots of
unity, and we expect that this holds whenever eq. \req{Alog} yields Bloch group 
elements of finite order.

For each Dynkin diagram of ADET type, the analysis of eq. \req{Alog} uses a 
family of polynomials which are linked by linear and quadratic recursion 
relations. For A$_1$ one finds the Chebysheff polynomials. To study the general
case one needs the representation theory of simple Lie groups and their
quantum group deformations. Relations to quivers and Ocneanu's essential 
paths on the Dynkin diagrams \cite{O99} are likely, but will not be explored 
here. Many different mathematical themes appear, from the elementary ones
central to this article up to rather advanced issues. The explanations
will take the mathematical reader straight into quantum field theory and
the physicist into some interesting areas of algebra. For both, parts of the  
article will be too well known to merit much attention, but when one
addresses a mixed audience this is unavoidable. More serious is the fact
that most of the obvious questions will remain open -- the end 
of this article is dictated by the present status of research and of my
understanding, not by any intrinsic logic. I hope that some readers will
do better. 

\subsubsection*{Acknowledgments}
I wish to thank Edward Frenkel, Herbert Gangl and Don Zagier for extensive
and helpful discussions which were essential for this article.

\section{Free and conformally invariant quantum field theories} 

The only quantum field theories which are easily understood from a
mathematical point of view are the free ones and (some of) the conformally
invariant ones. The intersection of these two families yields the theory
of massless fermions in two dimensions, which we will study as our basic
example. Mathematically this CFT is very
simple, but it allows to understand many general features of quantum field
theory, in particular the operator product expansion. The theory of free
massive fermions can be considered as a perturbation, and we will use it 
to consider basic features of perturbation theory.

We first show how to quantize arbitrary theories of free fermions.  Aspects
of conformal invariance will be introduced later. Free means linear, and
all one needs is a slight generalization of the quantization of the harmonic
oscillator. Recall that the phase space $V$ of a classical harmonic oscillator 
is a finite dimensional vector space with a non-degenerate anti-symmetric  
bilinear form. For definiteness one should consider a two-dimensional $V$  
with coordinates $x,p$ and a bilinear form obtained from $\{x,p\}=1$, but the
step to quantum field theory is easier when one uses a more abstract language.

The abelian group $V$ acts on the phase space (on itself) by translation,
and one still wants to have such an action on a space of states after 
quantization. Thus one constructs a Hilbert space $H$ on which $V$ is 
projectively represented. More precisely, $H$ carries a representation of
an extension of $V$ by the complex numbers of modulus one, and this
extension is given by the bilinear form.

In a more physical way the same procedure can be described as follows.
The linear functions on $V$ are observables of the classical theory.
One wants them to become observables of the quantized theory, too. The 
commutators of these observables yield the Heisenberg Lie algebra which is
given by the bilinear form, up to a factor of $i$. More precisely the vector
space of this Lie algebra is $\C\oplus V^\ast_\C$, where $V^\ast$ is the
dual of $V$ and $V^\ast_\C$ its complexification. This is just the
infinitesimal version of the previous description, since the linear
functions on $V$ generate translations in the Hamilton formalism and the
non-degenerate bilinear form yields a natural isomorphism between $V$ and 
$V^\ast$. In the standard two-dimensional example the observables satisfy the
Heisenberg Lie algebra given by $[x,p]=i$, translations of $x$ by $a$ are 
given by $\exp(-iap)$ and translations of $p$ by $\exp(iax)$. The extra
$i$ in $[x,p]=i$ compared to $\{x,p\}=1$ is necessary, since hermitean
$x,p$ yield anti-hermitean $[x,p]$. In the fermionic case it is unnecessary,
which saves us some trouble. 

For fermionic theories nothing much changes, the vector space $V$ just has
a symmetric bilinear from instead of an anti-symmetric one. Thus one works
with a super-Lie analogue of the Heisenberg algebra. Moreover only
the even polynomials in $V$ correspond to observables, so one has to
introduce a $\Z_2$ grading.  
 
Now we can start to construct the simplest quantum field theory. We will need
some care and elementary distribution theory to handle an infinite dimensional
$V$, but the main steps are just as before. Let $V$ be a real vector space
with non-degenerate symmetric bilinear form $\{,\}$. We regard $V$ as the odd 
part of a $\Z_2$-graded vector space $\R \oplus V$. 
The form defines a super-Lie algebra on $\R\oplus V$, with center $\R$,
and also on the complexification $\C\oplus V_{\C}$ of $\R\oplus V$.
We are interested in representations of this super-Lie algebra for
which $1\in\R$ is represented by the identity.

Let $V_{\C} = V_+ \oplus V_-$ be a decomposition of the complexification
of $V$ into isotropic, complex conjugate subspaces. Then we have
a natural super-Lie algebra representation 
$\hat\psi: V_{\C} \rightarrow End(H)$ with 
$$H=\Lambda V_-=\oplus_n \Lambda^n V_-,$$ 
such that $\hat\psi(v_+)$ annihilates $\Lambda^0 V_-=\C$ for $v_+\in V_+$,
whereas for $v_-\in V_-$, $w\in H$ one has 
$$\hat\psi(v_-)w=v_-\wedge w.$$
One finds that $V_+$ acts as a super-derivation, induced by  
$$\hat\psi(v_+)v_-=\{v_+,v_-\}.$$
The induced hermitean from on $H$ is not necessarily positive definite, but in 
our context we do not need $H$ to be a Hilbert space, since only direct sums
of finite dimensional spaces will occur. Theories with a positive definite form
are called unitary.

Let $\R\oplus V^1$, $\R\oplus V^2$ be two superalgebras of the type
just discussed, with representations on $H^i$, $i=1,2$.
When $V=V^1\oplus V^2$ and $\{v^1,v^2\}=0$ for $v^i\in V^i$,
then $\R\oplus V$ has a natural representation on $H^1\otimes H^2$.
For representations obtained from isotropic subspaces $V^i_-$
there is indeed a natural isomorphism $\Lambda (V^1_-\oplus V^2_-)\simeq
\Lambda V^1_- \otimes \Lambda V^1_-$.

The enveloping algebra of $\R\oplus V$ is the Clifford algebra given
by $V$ and $\{,\}$. When the dimension of $V$ is finite, our construction
demands that it is even, since $\dim(V)=2\dim(V_+)$. In this case it is
well known that all irreducible representations are isomorphic.
When $\dim(V)$ is odd, recall that we want a $\Z_2$ graded representation.
Let the $\Z_2$ grading of $H$ be given by 
$H=H_b\oplus H_f$ with an operator $\cal F$ which acts as $+1$ on $H_b$ and as
$-1$ on $H_f$. Thus we want a representation of $\R\oplus V \oplus
\langle {\cal F}\rangle$, which brings us back to the even case. 

But let us return to $V_{\C} = V_+ \oplus V_-$. Let $1^\ast\in H^\ast$ be
the projection to $\Lambda^0 V_-=\C$. The vectors $1\in H$ and
$1^\ast\in H^\ast$ are called vacuum vectors, and the vacuum expectation value
of an operator $O\in End(H)$ is denoted by $\langle O\rangle =1^\ast\,O\,1$. 
For odd $n$ and $v_i\in V$ we have
$\langle  \hat\psi(v_1) \hat\psi(v_2)\ldots \hat\psi(v_n)\rangle=0$, 
since $\hat\psi(v_i)\Lambda^n V_-
\subset \Lambda^{n-1} V_-\oplus \Lambda^{n+1} V_-$. For even $n$ we have

\beq{Wick}
\langle \hat\psi(v_1) \hat\psi(v_2)\ldots\hat\psi(v_n)\rangle = Pf(M),
\eeq
where $Pf$ is the Pfaffian and the anti-symmetric $n\times n$
matrix $M$ has entries
$M_{ij}=\langle \hat\psi(v_i)\hat\psi(v_j)\rangle$ for $i\neq j$ 
(Wick's thorem). We have

\beq{vv}
\langle \hat\psi(v_i)\hat\psi(v_j)\rangle = \{v_iv^-_j\},
\eeq
where $v^-_j$ is the projection of $v_j$ to $V^-$.

For the harmonic oscillator, phase space can be considered as the space
of initial conditions for the equations of motion, or more invariantly as
the space of solutions of these equations. The latter are linear
ordinary differential equations. In free quantum field theory, $V$ is 
the solution space of some linear partial differential equations.
We shall work in two dimensions, with time coordinate $t\in\R$ and space 
coordinate $x$. Often one considers $x\in\R$, but we shall concentrate on
the case of the unit circle $S^1$ with angle $x$.

We study the two-dimensional Dirac equation for a real two-component spinor 
$\psi=(\psi_R,\psi_L)$,
\begin{eqnarray}\la{Dirac} 
(\partial_x-\partial_t)\psi_R(x,t)&=\mu\psi_L(x,t)\nonumber\\
(\partial_x+\partial_t)\psi_L(x,t)&=\mu\psi_R(x,t),
\end{eqnarray}
with the symmetric bilinear form

$$(\psi^1,\psi^2)=\int_{t=t_0}(\psi_R^1\psi_R^2+\psi_L^1\psi_L^2)dx.$$
Note that this form is independent of the choice of $t_0$.
When the mass $\mu$ is different from zero it introduces a length scale, but
for $\mu=0$ the equations are invariant under the conformal transformations

$$(x-t,x+t)\mapsto (f(x-t),g(x+t)),$$ 
where $f,g$ are orientation preserving
diffeomorphisms of $S^1$. Later we will come back to the case $\mu\not=0$,
but in this section we will consider the conformally invariant theory.

For $\mu=0$ the equations for $\psi_R$ and $\psi_L$ decouple and can be treated
separately. Moreover, they can be solved trivially in terms of the value of 
$\psi$ at fixed time. Thus we can forget about differential equations and
regard the case where $V$ is the vector space of square
integrable functions on $S^1$, with the standard bilinear form given by the
measure $dx$. The spectrum of the rotation
generator allows to decompose $V_{\C}$ into a positive part $V_+$ spanned by
$\exp(imx)$ with $m>0$, a negative part $V_-$ with $m<0$, and the constants, or
zero modes, which form a one-dimensional vector space $V_0$. This is the R case
considered by Ramond. Since the bilinear form is invariant under rotations,
$V_+$ and $V_-$ are isotropic. The constants are orthogonal to $V_+,V_-$, such 
that $\R\oplus V$ has representations on $\Lambda^n V_-\otimes H_0$, where 
$H_0$ is the standard two-dimensional representation space of  
the quaternion algebra given by $\R\oplus V_0\oplus\langle{\cal F}\rangle$.
A basis of $\Lambda^n V_-$ or equivalently of $\Lambda^n V_+$ corresponds 
to the partitions of natural numbers into $n$ distinct summands $m>0$,
which yields one of the partition functions mentioned in the introduction.

The complications with the zero modes can be avoided when one
considers the anti-periodic functions on the double cover of
$S^1$ instead, in other
words the sections of the M\"obius bundle. This is the NS case considered
by Neveu and Schwarz. All fermionic CFTs have NS and R sectors, so it is
useful to introduce them together. 
In the NS case, $V_+$ is spanned by $\exp(irx)$, where $r=1/2,3/2,\ldots$.
A basis of $\Lambda^n V_+$ corresponds to the partitions of natural numbers
into $n$ distinct odd summands $2r$.

The NS case arises naturally when we take into account conformal invariance.
The scalar product $\int f(x)g(x)dx$ of two functions is no longer
natural, since the differential $dx$ on $S^1$ changes under diffeomorphisms.
Instead we have to factorize
$$\psi^1(x)\psi^2(x)dx=\left(\psi^1(x)\sqrt{dx}\right)
\left(\psi^2(x)\sqrt{dx}\right).$$
Thus $V$ should be regarded as a vector space of half-differentials,
in other words of sections of a squareroot of the cotangent bundle.
Two different squareroots exist, which correspond to the NS and R cases.
Considering $S^1$ as the boundary of the complex unit disc we write
$z=\exp(ix)$. The cotangent bundle on the disc has a unique squareroot 
with a section $\sqrt{dz}$. On the boundary it reduces to the 
complexification of
the Moebius bundle with section $\sqrt{dz} = \exp(ix/2)\sqrt{dx}$.
Since $\exp(ix/2)$ is anti-periodic, the NS case is in a sense more 
basic than the R case.

Anticipating some future simplifications we write
$$\hat\psi \sqrt{2\pi i dx} = \psi \sqrt{dz}.$$
This is just a change of normalization by $\exp(ix/2)/\sqrt{2\pi}$.

The mathematical reader may wonder what all of this has to do with
quantum field theory, since it is just very conventional mathematics,
but now comes the essential step.
We can consider $\psi$ as an operator valued distribution, which to
a function $v$ on $S^1$ associates the operator $\psi(v)$ on $H$.
Note that $\psi(v)\psi(v')=-\psi(v')\psi(v)$ when the supports of $v,v'$
have empty intersection. In physics terminology, such operator valued 
distributions are called fermionic local fields on $S^1$.  
In general a local field theory on $S^1$ provides a $\Z_2$-graded vector 
space of fields $F=F_b\oplus F_f$, with degree $\eta(\phi)=0$ for $\phi\in F_b$
and $\eta(\phi)=1$ for $\phi\in F_f$ such that
$$\phi(v)\chi(v')= (-)^{\eta(\phi)\eta(\chi)}\chi(v')\phi(v)$$
when the test functions $v,v':S^1\rightarrow \C$ have non-intersecting
support. The space $F_b$ contains the bosonic fields and $F_f$ the
fermionic ones.

In this section we only need fields on $S^1$, but let us
indicate what happens in more general contexts. The functions $v,v'$
become test functions on a spacetime with causal structure
and the equation $\phi(v)\chi(v')=\pm\chi(v')\phi(v)$ applies, with appropriate
sign, when the supports of $v,v'$ are causally independent.

The simple case we consider is called the quantum field theory of
a free fermion and the distribution $\psi$ is a free fermion field. The
vector $1\in H$ is called the vacuum, and the vacuum expectation value
$\langle \psi(v_1)\psi(v_2)\ldots \psi(v_n)\rangle$
can be considered as the value of a distribution on the $n$-fold Cartesian
product of the circle, evaluated at $v_1\otimes\ldots\otimes v_n$.
This distribution is written $\langle \psi\ldots\psi\rangle$.
By Wick's theorem \req{Wick}, it is sufficient to calculate  
$\langle\psi\psi\rangle$. By eq. \req{vv} we have

$$\langle \hat\psi(v)\hat\psi(v')\rangle= \int v(x)v'_-(x)dx$$
where
$$v'_-(x)=\sum_{r>0} \exp(-irx)\,\int v'(y)\exp(iry)\ \frac{dy}{2\pi}.$$
Thus the distribution $\langle\hat\psi\hat\psi\rangle$ is the
$\epsilon\rightarrow +0$ limit of the function 
$$ \frac{1}{2\pi}\sum_{r>0} \exp(-irx)\exp(iry) = 
\frac{1}{2\pi}\exp(ix/2)\exp(iy/2)\,\frac{1}{\exp(ix)-\exp(iy)}$$
where $\Im x =0$, $\Im y = \epsilon$,
such that the absolute value of $\exp(iy)$ approaches the one
of $\exp(ix)$ from below. In terms of $\psi = \sqrt{2\pi} 
\exp(-ix/2)\hat\psi$, we see that the distribution 
$\langle\psi\psi\rangle$ is the limiting value of the function
$$\langle\psi\psi\rangle(z,w) = (z-w)^{-1}$$
where $z=\exp(ix)$, $w=\exp(iy)$, and $|z|$ approaches $|w|$ from above. 
More generally, for even $n$ the distribution $\langle\psi\ldots\psi\rangle$ on 
the $n$-fold Cartesian product of the circle is a limiting value of a function
on $\C^n$ given by
$$\langle\psi\ldots\psi\rangle(z_1,\ldots,z_n)=Pf(M)$$
with $M_{ij}=(z_i-z_j)^{-1}$ off the diagonal. Such functions are
called $n$-point functions. By convention, they are written in the form
$\langle\psi(z_1)\ldots\psi(z_n)\rangle$. In the distributional limit the 
$|z_k|$ all tend to 1 and the limit is taken along a path such that 
$|z_i|>|z_j|$ for $i<j$. The only singularities of the $n$-point function occur 
at the partial diagonals $z_i=z_j$. For fermionic fields, as in the
present case, the $n$-point functions are anti-symmetric in all
variables, which we express in the form $\psi(z)\psi(w)=-\psi(w)\psi(z)$.

In general, quantum field theories can be formulated either in terms of 
operator valued distributions or of $n$-point functions. We mainly will
use the latter formulation, which is called euclidean quantum field theory.
So far we considered a single field $\psi$, but in general one needs 
bosonic and fermionic fields in a vector space $F=F_b\oplus F_f$. A rather
trivial but important element is the identity field $I\in F_b$ which does
not depend on $z$ and satisfies
$$\langle I\phi^1(z_1)\ldots\phi^n(z_n)\rangle =
\langle\phi^1(z_1)\ldots\phi^n(z_n)\rangle$$
for all $\phi^k\in F$. Let $T(F)=\oplus_n T^n(F)$ be the
tensor algebra over $F$. Then the $n$-point functions map $ T^n(F)$ to
the space of functions on $\C^n$ minus the partial diagonals, and the
map factors through the projection from $T(F)$ to $SF_b\otimes \Lambda F_f$.

For $N$ quantum field theories with field spaces $F^k$, $k=1,\ldots,N$, one
has a natural tensor product with field space $\otimes F^k$. The $n$-point 
functions of the product theory are the products of the $n$-point functions
of the factors. There is a natural embedding $F^i\rightarrow \otimes F^k$,
$i=1,\ldots,N$, since one can identify $I\otimes\phi$ with $\phi$. For $N$
free fermion theories the product has  fields $\psi^k$, $k=1,\ldots,N$, with 
$n$-point functions
\beq{vevN}
\langle\psi^{k_1}(z_1)\ldots\psi^{k_n}(z_n)\rangle=Pf(M),
\eeq
for even $n$, where now
$$M_{ij}=\delta_{k_i,k_j}(z_i-z_j)^{-1}.$$
Equivalently we could generalize our simplest example such that $V$ is given
by maps from $S^1$ to $\R^N$, with standard bilinear form. Then every element
$a\in \R^N$ yields one local quantum field $\psi_a\in F_f$ which maps functions
$f:S^1 \rightarrow \R$ to $\psi_a(f)=\psi(fa)$. The $\psi^k$ the correspond
to the standard basis of $\R^N$.

We abstract certain features from our examples, which will be valid
in general. There is a vector $1\in H$ and a vector $1^\ast$ in its
dual such that the $n$-point functions
$\langle\phi^1(z_1)\ldots\phi^n(z_n)\rangle =
1^\ast \phi^1(z_1)\ldots\phi^n(z_n)1$ are translationally invariant,
$$\langle\phi^1(z_1+z_0)\ldots\phi^n(z_n+z_0)\rangle
=\langle\phi^1(z_1)\ldots\phi^n(z_n)\rangle.$$
These functions are real analytic away from the partial diagonals.
When all $n$-point functions involving $\phi\in F$ vanish, then $\phi=0$.
Thus fields can be determined by their $n$-point functions.

A field $\phi$ is called holomorphic when all $n$-point functions
$\langle \phi(z)\ldots\rangle$ depend meromorphically on $z$, as in our   
basic example. For each pair of fields $\phi, \phi^1$ with holomorphic $\phi$
we construct a family of fields $N_m(\phi, \phi^1)$ whose
$n$-point functions are given by the Laurent expansion of $(n+1)$-point
functions involving $\phi, \phi^1$, i.e.
\beq{OPE}
\langle\phi(z)\phi^1(z_1)\ldots\phi^n(z_n)\rangle =
\sum_m (z-z_1)^m
\langle N_m(\phi, \phi^1)(z_1)\phi^2(z_2)\ldots\phi^n(z_n)\rangle
\eeq
in the sense of a Laurent expansion in $z$ around the possible pole
at $z=z_1$.
Note that this equation is assumed to be valid for all choices of the
$\phi^k$. Symbolically it is written in the form

$$\phi(z)\phi^1(w)= \sum_m (z-w)^m N_m(\phi, \phi^1)(w).$$

This is the simplest case of the operator product expansion (OPE) of
local fields. The field $N_0(\phi, \phi^1)$ is called the normal
ordered product of $\phi, \phi^1$. The OPE also can be defined when $\phi$
is not holomorphic, but then the functions $(z-w)^m$ are replaced by 
non-universal sets of real analytic functions.

The OPE is compatible with the $\Z_2$-grading. In particular, for
$\phi,\phi^1\in F_f$ one has $N_m(\phi,\phi^1)\in F_b$ for all $m$.
For the free fermion, $N_{-1}(\psi,\psi)=I$.

In our basic example the $n$-point functions are homogeneous with
respect to linear transformations of $\C$. More generally, conformal
invariance yields a grading $h$ of the vector space of holomorphic
fields, such that their $n$-point functions satisfy
\beq{holgr}
\langle \phi^1(az_1)\ldots \phi^n(az_n)\rangle =
a^{-h_\Sigma}\langle \phi^1(z_1)\ldots \phi^n(z_n)\rangle, 
\eeq
where
$$h_\Sigma=\sum_{k=1}^n  h(\phi^k).$$
One easily sees that
\beq{hN}
h\left(N_m(\phi^1,\phi^2)\right)=m+h(\phi^1)+h(\phi^2).
\eeq
In particular, $h(I)=0$, and $h(\psi)=1/2$ for free fermion fields.
More generally, bosonic holomorphic fields have integral and fermionic ones
half-integral conformal dimension. When one lets $a$ vary over the unit
circle, one sees thate $n$-point functions involving an odd number of fermionic 
fields have to vanish.
 
Now we can define holomorphic field theories. There is a vector space
$F$ with a grading $F=\sum_{h\in\N/2}\,F(h)$ such that $F(0)=\C$. Let 
$\oplus_{h\in\N}\,F(h)=F_b$ and $\oplus_{h\in\N+1/2}\,F(h)=F_f$.
There is an $n$-point function map from $SF_b\otimes \Lambda F_f$ to the
meromorphic functions which are holomorphic away from the partial diagonals,
such that these functions are translationally invariant and satisfy the
scaling relation \req{holgr}. With respect to these functions, $F$ must be 
closed under the OPE. No field must have identically vanishing $n$-point
functions. Finally, for sufficently large $N$ and all $h$
$$\dim\, F(h)\leq \dim\,F^N(h),$$
where $F^N$ is the field space of the theory of $N$ free fermions.

A large class of such theories can be constructed by taking the OPE closure
of some subspace of $F^N$ and factoring out those fields which have 
identically vanishing $n$-point functions. One example is $F^N_b$, but we will
consider several others. Some further constructions will be considered below.  

Scale invariant non-holomorphic theories can be defined in the same way.
First let us construct examples. Take a holomorphic theory with field space
$F_{hol}$, and the complex conjugate of the $n$-point functions. This is a
theory of anti-holomorphic fields, with a field space $F_{\overline{hol}}$ 
anti-linearly isomorphic to $F_{hol}$. The tensor product of the two theories
has a field space $F_{hol}\otimes F_{\overline{hol}}$. There are subspaces
$F_{hol}$ and $F_{\overline{hol}}$, but all the remaining fields are neither
holomorphic nor anti-holomorphic.

Let us generalize this example. There must be a double grading $(h,\bar h)$
of $F$, such that
\beq{scaling}
\langle \phi^1(az_1)\ldots \phi^n(az_n)\rangle =
a^{-h_\Sigma}\bar a^{-\bar h_\Sigma}
\langle \phi^1(z_1)\ldots \phi^n(z_n)\rangle,
\eeq
where
\begin{eqnarray*}
h_\Sigma&=&\sum_{k=1}^n  h(\phi^k)\\
\bar h_\Sigma&=&\sum_{k=1}^n  \bar h(\phi^k).
\end{eqnarray*}
The sum $h(\phi)+\bar h(\phi)$ is called the scaling dimension of $\phi$,
and $h(\phi)-\bar h(\phi)$ is called its conformal spin. For holomorphic $\phi$
one has $\bar h(\phi)=0$. When we rewrite
$$a^{h_\Sigma}\bar a^{\bar h_\Sigma}=
|a|^{h_\Sigma+\bar h_\Sigma}(a/|a|)^{h_\Sigma-\bar h_\Sigma}$$
in eq. \req{scaling}, we see that one can admit arbitrary real scaling
dimensions, as long as the conformal spins are integral for bosonic fields and 
half-integral for fermionic ones. For non-holomorphic fields, the conformal 
dimensions  $(h,\bar h)$ may be negative. One even could admit complex scaling 
dimensions, but we will avoid that. 

One demands that the $n$-point functions are real analytic away from the 
the partial diagonals, and that $F$ is closed under the OPE. The latter is 
slightly more complicated to define than in the holomorphic case. One best
works with function germs, but we will not need the details. To control
the growth of the dimensions, one works with the filtered subspaces
$\oplus_{h+\bar h\leq d}F(h,\bar h)$ and demands that for some $N$ and 
suffiently  large $d$ the dimension of these spaces is smaller than for the
theory $F^N\otimes \bar F^N$ of $N$ holomorphic and $N$ anti-holomorphic free 
fermions.

In many cases, $F$ has a real structure such that the $n$-point functions
for real fields at real $z^i$ take real values. For example, our free
fermion fields $\psi$ are real. The complex conjugate theory of such
a theory has a real structure, too. Often the anti-holomorphic free
fermion field is called $\bar\psi$. This may be confusing, since both
fields are real and linearly independent. The OPE of real fields yields
real fields. 

Using the OPE to deduce properties of
the $n$-point functions from those of the 2-point function, one can read
off from the scaling behaviour of the latter that for $z\rightarrow\infty$
\beq{infty}
|\langle\phi(z)\ldots\rangle|=O\left(|z|^{-h(\phi)-\bar h(\phi)-\tilde
d(\phi)}\right),
\eeq
where $\tilde d(\phi)$ is the minimal scaling dimension $h(\tilde \phi)+
\bar h(\tilde\phi)$ of all fields $\tilde\phi$ for which
$\langle\phi\tilde\phi\rangle\neq 0$. Thus $n$-point fuctions have good
behaviour at infinity.

Consider a holomorphic field theory for which $F$ is the OPE closure of 
$F(1)$. Let $F(1)$ be spanned by the fields
$J_a$, $a$ in some finite index set. Their OPE must have the form
\beq{currentOPE}
J_a(z) J_b(w)=\frac{d_{ab}}{(z-w)^2}+
\sum_c\frac{f_{abc}}{z-w} J_c(w) +\quad\mbox{regular terms}.
\eeq
Given the structure constants $d_{ab}$ and $f_{abc}$, this is sufficient
to calculate all $n$-point functions of the theory, since by induction
one knows all singular terms of these meromorphic functions, and also 
their behaviour at infinity. Using the OPE for the four-point functions
in the various possible ways, one sees that the $f_{abc}$ must be the
structure constants of a Lie algebra, and $d_{ab}$ must yield a non-degenerate 
invariant bilinear form. The $J_a$ are called currents of this Lie algebra. 
When the latter is semi-simple, we shall see that the Fourier components of 
its currents generate the corresponding affine Kac-Moody algebra.

As an example, consider the theory of $N$ free fermions, with $n$-point
functions given by eq. \req{vevN}. Here $F(1)$ is spanned by the fields
\beq{JN}
J_{ij}=N_0(\psi^i\psi^j),
\eeq
$i,j=1,\ldots,N$, $i<j$. One easily calculates that their OPE has the form  
\req{currentOPE} where $a,b,c$ stand for pairs $(ij)$, the invariant bilinear
form is given by $d_{ab}=\delta_{ab}$, and the $f_{abc}$ are the standard
structure constants of the Lie algebra $so(N)$. 

More generally, we are particularly interested in the eq. \req{currentOPE}
when the $f_{abc}$ are the normalised structure constants of a simple
Lie algebra $X$ and 
\beq{dk}
d_{ab}=k\delta_{ab} 
\eeq
for $k=1,2,\ldots$. We call the corresponding space of fields $F_X^k$. We just
saw that $F_{so(N)}^1\subset F^N$, where $F^N$ is the field space for $N$ free
fermions. One even can show that
$$F_{so(N)}^1=F^N_b.$$

For fixed fields $\phi^i$, $i=1,\ldots,n$ and $z_i$ away from the partial
diagonals and 0, the $n$-point function values
$$\langle \phi^1(z_1)\ldots\phi^n(z_n)\phi(0)\rangle$$
can be interpreted as a linear form acting on $\phi\in F$. In physics
it is traditional to distinguish the space $H$ spanned by the $\phi(0)1$
from $F$ and to call the map $\phi\mapsto \phi(0)1$ field-state identity.
Indeed, in the free fermion case we first constructed $H$ and then $F$.
It is easy to show that one has an isomorphism, since the $n$-point
functions of $\phi$ must not vanish identically and are translationally
invariant. Sometimes one wants to be $H$ a completion of $F$, but we do not
need that. If you want, put $H=F$.
 
For holomorphic $\phi$ the maps $N_m(\phi,.):
F\rightarrow F$ yield operators on $H$. These are the Fourier
components $\phi_{m+h}$ of $\phi$. They increase the degree of a vector by
$m+h$, where $h=h(\phi)$. In the literature the sign of
the index often is inverted, such that this operator is denoted
by $\phi_{-m-h}$, but this seems unnecessarily confusing. In any case one
obtains this operator by
evaluating the field $\phi$ on $z^{-m}$, with respect to the
measure $dx/(2\pi)=dz/(2\pi iz)$. In our convention, the field-state 
identity is given by $\phi\mapsto \phi_{h(\phi)}1$.

For holomorphic fields we now can move from the language of euclidean field
theory to the one of operator valued distributions and back. In euclidean
field theory everything is commutative, but the singularities of an
$n$-point function yield non-vanishing commutators for the limiting 
distribution $\langle \phi^1\ldots\phi^n\rangle$ on the $n$-th Cartesian power
of the unit circle. Recall that the limit to $|z^k|=1$ for $k=1,\ldots,n$ has
to be taken from the domain with $|z_1|>\ldots>|z_n|$. Evaluating the
distribution on monomials in the $z_i$ yields the matrix elements
$$\langle \phi^1_{k_1}\ldots\phi^n_{k_n}\rangle
= \langle \phi^1\ldots\phi^n\rangle\ \left( \prod_i z_i^{-k_i+h_i}\right),  $$
where $h_i=h(\phi^i)$. By Cauchy's theorem this expression can be evaluated
before the limit is taken, by integrating along circles with ordered radii
$|z_1|,\ldots,|z_n|$. Thus one obtains the Laurent expansion of the
$n$-point function in the domain $|z_1|>\ldots>|z_n|$ by inserting

$$\phi^i(z) = \sum_m \phi^i_m z^{m-h_i},$$
keeping the order of the fields.
When one changes the order of the operators $\phi^k_m$, the vacuum
expectation value is given by integrating the same $n$-point function along
differently ordered circles. By Cauchy's theorem, the difference of the two 
expressions only depends on the singularities of the function, in other words
on the singular part of the OPE.

In a holomorphic tensor product theory with $F=F^1\otimes F^2$ let
$\phi^i\in F^i$, $i=1,2$. The OPE for $\phi^1(z)\phi^2(w)$ has no
singularities, such that $[\phi^1_m,\phi^2_n]=0$ for all $m,n$.
Given an OPE closed subspace $G\subset F$ of an arbitray holomorphic
theory, one can find a maximal complement $G'\subset F$, such that one
has an embedding $G\otimes G'\subset F$.  Note the $G'$ is
closed under the OPE and yields a new holomorphic field theory. A field $\chi$
belongs to $G'$, iff its OPE with arbitrary fields in $F$ is non-singular,
or equivalently, iff the Fourier components commute. This procedure is
very important for the construction of new theories and will be used later.

For currents $J_a$ with an OPE \req{currentOPE} Cauchy's theorem yields 
$$[J_{am}J_{bn}]=nd_{ab}\delta_{m+n,0}+f_{abc}J_{c,m+n}.$$
Obviously the $J_{a0}$ span a finite dimensional Lie algebra with
structure constants $f_{abc}$. When this Lie algebra is simple, the $J_{an}$ 
and 1 span the corresponding affine Kac-Moody Lie algebra. We shall use 
some properties of this algebra without derivation. A standard reference
is \cite{VK90}. 

In the case of the tensor product $F^N$ of $N$ free fermion theories,
the $so(N)^{(1)}$ Kac-Moody algebra is represented on $H_{NS}$ and $H_R$. The
$\Z_2$ grading of $H_{NS}$ given by $F^N=F^N_b\oplus F^N_f$ yields a 
decomposition into two subrepresentations, which turn out to be irreducible.
The representation on $H_R$ will be considered in more detail below. The
 $\Z_2$ grading again yields two irreducible subrepresentations, which are
isomorphic for odd $N$. In any case one obtains all irreducible level 1 
representations of $so(N)^{(1)}$.

We now had a glimpse of the interesting applications of the operator
product expansion and more will come later. I think that the OPE
is one of the major contributions of 20th century physics to mathematics, but
still one of the major stumbling blocks when mathematicians try to learn
quantum field theory. Indeed on first sight it is hard to believe that it
makes sense. Take a small vector space $F$ and write any set of candidates
for its $n$-point functions. Then take the OPE closure. On first sight it is
hard to believe that this can yield a  managable space of local fields. For
holomorphic theories this is now well understood, but that it happens in many
more interesting examples is one of the miracles of quantum field theory.

The particular OPE for $\phi(z)I$ yields derivative fields. For holomorphic
$\phi$ one has $\partial^m\phi=N_m(\phi,I)/m!$, such that
$$\langle(\partial\phi)(z)\chi(w)\rangle = \partial_z
\langle\phi(z)\chi(w)\rangle.$$
Note that $h(\partial\phi)=h(\phi)+1$. The generalization to arbitrary
fields and anti-holomorphic derivatives is immediate. Eq. \req{OPE} yields
$$\partial N_m(\phi,\chi)= N_m(\partial\phi,\chi)+N_m(\phi,\partial\chi).$$ 

In conformally invariant theories one demands the existence of Virasoro fields.
They  are defined by
$$N_{-1}(T,\phi)=\partial\phi$$
for all $\phi\in F$. Obviously one needs $h(T)=2$, $\bar h(T)=0$. To check that 
the definition makes sense, let $N_{-1}(T,\phi_i)=\partial\phi_i$ for $i=1,2$,
and let $\phi_1$ be holomorphic.
By acting with $\oint T(z)dz$ on the OPE and applying Cauchy's theorem one
finds indeed that $N_{-1}(T,\phi)=\partial\phi$ for $\phi=N_m(\phi_1,\phi_2)$
and any $m$.
The map $\phi\mapsto N_{-2}(T,\phi)$ is grade preserving. In all examples
considered here it is diagonalizable and coincides with the grading,
$$N_{-2}(T,\phi)=h(\phi)\phi.$$
Indeed, by acting with $\oint zT(z)dz$ on the OPE and use of Cauchy's theorem, 
we see that $N_{-2}(T,\phi_i)=h_i\phi_i$ for $i=1,2$ implies
$N_{-2}(T,\phi)=(m+h_1+h_2)\phi$ for $\phi=N_m(\phi_1,\phi_2)$ and any $m$.
 
For historical reasons, the Fourier components of $T$ are called
$L_m$. Thus the grading operator is $L_0$. One always assumes that
the spectrum of $L_0$ is bounded from below. In physical terminology,
one considers highest weight representation, though `lowest weight'
would make more sense. States of lowest degree are often called ground states.

In a CFT with non-holomorphic fields we also have an
anti-holomorphic Virasoro field $\bar T$ with Fourier components $\bar L_m$,
such that $\bar L_0\phi=\bar h(\phi)\phi$. The scaling dimension is the
eigenvalue of $L_0+\bar L_0$. It describes the behaviour of the
$n$-point functions under the transformations $z\mapsto az$ with real positive
$a$. This corresponds to time translations of the theory on $S^1\times\R$,
up to an additive constant $-c/24$, which will be explained later.
The eigenvalues of $L_0-\bar L_0$ have to be integral and describe the
effect of transformations $z\mapsto az$ with $|a|=1$. The latter correspond
to rotations of $S^1$, i.e. to translations of the angle $x$.

For the free fermion field $\psi$ one finds 

$$\psi(z)\psi(w) = (z-w)^{-1}I + 2(z-w)T(w) + O((z-w)^3).$$
Equivalently, $T=N_1(\psi,\psi)/2=N_0(\partial\psi,\psi)/2$. 
For the OPE of $T$ one obtains

$$T(z)T(w) = \frac{1}{4} (z-w)^{-4}I +  2(z-w)^{-2}T(w)
+ (z-w)^{-1}\partial T(w) +O(1)$$
or more suggestively

$$T(z)T(w) = \frac{1}{4}(z-w)^{-4}I +  (z-w)^{-2}(T(z)+T(w)) +O(1).$$
 
If we write the Virasoro OPE in this symmetric way, no odd powers of
$z-w$ can occur. The term proportional to $(z-w)^{-2}$ is fixed by
$N_{-2}(T,T)=2T$. The term proportional to $(z-w)^{-4}$ has conformal
dimension 0, thus must be a constant. Thus the OPE of any Virasoro field
has the form

$$T(z)T(w) = \frac{c}{2} (z-w)^{-4}I +  (z-w)^{-2}(T(z)+T(w)) +O(1)$$
where $c$ is a constant, called the central charge of the theory. For the
free fermion we just found $c=1/2$. 

The tensor product of two CFTs with Virasoro fields $T^1,T^2$ is a CFT
with Virasoro field $T^1+T^2$. The central charges add up. When $F$ is a
holomorphic theory with
Virasoro field $T$ and $G\subset F$ a theory with Virasoro field $T^1$,
then the complementary subtheory $G'$ has a Virasoro field $T-T^1$.

Consider now the holomorphic field theory $F^k_X$ defined by eqs. 
\req{currentOPE} and \req{dk}. Recall that $F(1)$ is spanned by the currents 
$J_a$ and that $X$ itself is given by the commutators of the $J_{a0}$. The space
$F(2)$ of fields with conformal dimension 2 is spanned by the derivatives  
$\partial J_a$ and the bilinears $N_0(J_a,J_b)$. Only a one-dimensional subspace 
of $F(2)$ is invariant under $X$, namely the multiples of 
$\sum_a\, N_0(J_a,J_a)$. A particular multiple is a Virasoro field, with
central charge related to the Coxeter number $h(X)$ by
$$c(X,k)=\frac{k\,\dim(X)}{k+h(X)}.$$
The calculation is straightforward, but somewhat lengthy and we will no
consider it here. In the case of ADE algebras $X$ of rank $r(X)$ we have
$$\dim(X)-r(X)=r(X)h(X),$$
such that $c(X,1)=r(X)$. This is the only case we will consider in this article. 

Let $J_a$ with $a=1,\ldots, r(X)$ generate a maximal abelian subalgebra
of $X$. The OPE completion of their span has a Virasoro field
$$T=\sum_{a=1}^{r(X)}\,\frac{1}{2}\,N_0(J_a,J_a)$$
with central charge $r(X)$. This is the same as $c(X,1)$, and one obtains
the surprising result that the Virasoro field of this subtheory is
identical to the Virasoro field of $F^1_X$, and independent of the choice
of the maximal abelian subalgebra. This is an important special
feature of the ADE cases. One can express it in the form that the
complement of the subtheory is trivial. The calculation which yields
central charge $r(X)$ is an easy and instructive exercise.

Now consider $k$ copies of $F^1_X$ and the product theory on 
$(F^1_X)^{\otimes k}$. Let $J^i_a$, $i=1,\ldots,k$ be the currents of the 
factor theories. There is a natural embedding
$F^k_X\rightarrow (F^1_X)^{\otimes k}$, where the image of $F^k_X(1)$ is spanned
by the currents 
$$\hat J_a = \sum_{i=1}^k J^i_a.$$
Let $\tilde F^k_X$ be the complement 
of this subtheory. For its central charge one obtains immediately
\beq{ctilde}
\tilde c(X,k)=\frac{k(k-1)r(X)}{k+h(X)}.
\eeq

Let $F_{ab}$ (with $ab$ for abelian) be the subtheory generated by the 
$\hat J_a$ for $a=1,\ldots,r(X)$. Its central charge is $r(X)$ and the
central charge of its complement $F_{ab}'$ in $(F^1_X)^{\otimes k}$ is
$(k-1)r(X)$. The complement $\check F^k_X$ of $\tilde F^k_X$ in $F_{ab}'$ 
has central charge
\beq{ccheck}
\check c(X,k)=\frac{h(X)(k-1)r(X)}{k+h(X)}.
\eeq
The $\check F^k_X$ theory also can be described as the complement of
$F_{ab}$ in $F^k_X$. 

We now have constructed most of the examples of CFTs which we will need to
illustrate the relations to algebraic K-theory. The remaining examples are
minimal models. These are rational CFTs which are the OPE closures of
their respective Virasoro fields. In such a theory one has $F(1)=0$ and
$F(2)=\langle T\rangle$. 

Taking Fourier components and using Cauchy's theorem for the Virasoro OPE
yields the Virasoro algebra

$$[L_mL_n]=(n-m)L_{m+n}+\frac{c}{12}(n^3-n)\delta_{m+n,0}.$$
Minimal models can be described by the representation theory of this algebra,
but we shall follow the euclidean formulation, which provides instructive
exercises for newcomers in quantum field theory. Because of their OPE, the
residues of the $n$-point function for $n$ Virasoro fields are given by the 
$(n-1)$-point function. Moreover, the functions vanish at infinity, so they can 
be calculated inductively. For the two-point function one finds
$$\langle T(z)T(w)\rangle = {c\over 2} (z-w)^{-4}$$
for the three-point function
$$ \langle T(z)T(w)T(u)\rangle =
c((z-w)(w-u)(u-z))^{-2}$$
and for the four point functions
\begin{eqnarray*}
\langle T(z)T(w)T(u)T(v)\rangle =&  F(z,w,u,v)+F(z,u,w,v)+F(z,v,w,u)\\
&+ G(z,w,u,v)+G(z,u,w,v)+G(z,w,v,u) 
\end{eqnarray*}
where
$$ F(z,w,u,v) = \langle T(z)T(w)\rangle \langle T(u)T(v)\rangle$$
and
$$ G(z,w,u,v) = c((z-w)(w-u)(u-v)(v-z))^{-2}.$$
For a rapid calculation involving normal ordered products, let

$$T(z)T(w) = {c\over 2} (z-w)^{-4}I + (z-w)^{-2}\left(T(z)+T(w)\right) +
\Phi(w)+ O\left((z-w)^2\right).$$
This yields
$$\langle T(z)\Phi(w)\rangle = -2c(z-w)^{-6}$$
and
$$\langle \Phi(w)T(u)T(v)\rangle = \frac{c(c+2)}{2}\left((w-u)(w-v)\right)^{-4}
-4c(w-u)^{-3}(w-v)^{-3}(u-v)^{-2}.$$
Note that $\Phi = N_0(T,T)-\partial^2T/2$. With 
\beq{U}
U = N_0(T,T)-{3\over 10}\partial^2T
\eeq
one finds $\langle T(z)U(w)\rangle = 0$ and
$$ \langle U(w)T(u)T(v)\rangle = \frac{c}{2}\left(c+{22\over 5}\right)
((w-u)(w-v))^{-4}.$$
Field theories which are generated by a Virasoro field alone
are called minimal models. In a minimal model the space of fields
of degree at most 5 is spanned by $I$, $T$, $\partial T$, $\partial^2 T$,
$\partial^3 T$, $U$ and $\partial U$. These are the fields which can occur
in the singular part of the OPE for $T(w)U(z)$. For  $c=-22/5$ the preceding
formulas show, however, that this singular part cannot involve $I$, $T$  or its
derivatives. Thus the singular terms in the Laurent expansion of 
$\langle U(z)T(w_1)\cdots T(w_n)\rangle$ around $z=w_n$ are given by
$\langle U(z)T(w_1)\cdots T(w_{n-1}\rangle$. Since these functions go to zero
for large $z$, they vanish by induction in $n$.
Thus in the minimal model with $c=-22/5$ any $n$-point function
involving $U$ vanishes and we conclude that $U=0$. In a sense, this
means that $c=-22/5$ yields the simplest minimal model. 

The value $-22/5$ looks a bit strange, so let us explain the pattern of
central charges in minimal models. Minimal models are classified by
pairs $p,q\in\N$ such that $1<p<q$, and $p,q$ have no common divisor.
Their central charge is $c=1-6(p-q)^2/pq$. The pair $(2,3)$
yields a trivial theory with $c=0$ and $H=\C$. The pair $(2,5)$ yields
$c=-22/5$, whereas the pair $(3,4)$ yields the bosonic part of the free fermion 
theory and $c=1/2$.

For perturbation theory we will need more complicated non-holomorphic fields
than those considered so far, but for the purpose of this article we need 
little more to know about them than their conformal dimensions. These are
constrained by the algebra of the holomorphic fields in a way which is best 
discussed in terms of representation theory. According to the field-state 
correspondence, all fields should correspond to vectors $v\in H$. For arbitrary
homogeneous vectors $v_1\in H^\ast$, $v_2\in H$ and holomorphic fields $\phi^k$
we can study the generalized $n$-point functions
$\langle v_1|\phi^1(z_1)\ldots\phi^n(z_n)|v_n\rangle$. The Laurent expansions
around the partial diagonals are given by the OPE of the holomorphic fields, 
which can be read off from the ordinary $n$-point functions, as discussed 
above. When all fields $\phi^k$ are bosonic or when $v_1,v_2$ belong
to the NS sector, the generalized $n$-point functions have the same analytic 
properties as the ordinary ones, except for the fact that poles also occur at 
$z_i=0$ and $z_i=\infty$. The order of the pole at $z_i=0$ is bounded by
$h(\phi^i)+c_2$, where $c_2$ is a constant given by the holomorphic degree of 
$v_2$ and independent of $n$ and of the choice of fields. Similarly, the 
pole order at $\infty$ is bounded by $h(\phi^i)+c_1$, with an analogous
constant $c_1$.

When $v_1,v_2$ belong to the R sector and $\psi\in F_f$, then
$\langle v_1|\psi(z)\ldots\rangle$ changes sign when $z$ moves around the
origin. To recover the standard analytic behaviour of the generalized
$n$-point functions we have to multiply 
$\langle v_1|\phi^1(z_1)\ldots\phi^n(z_n)|v_n\rangle$ by the product of all
$\sqrt{z_k}$ for which $\psi_k\in F_f$. When $v_1$ belongs to the NS sector
and $v_2$ to the R sector, the functions have to vanish. 

Given an OPE of holomorphic fields $F$, we can constrain the
vectors $v\in H$, or rather the vectors $v_1\otimes v_2\in H^\ast\otimes H$ 
by abstracting from these properties. Let ${\cal F}(n)$ be the space of 
meromorphic functions on $\C^n$ with poles only on the partial diagonals, 0 
and infinity. A NS state on the OPE is a map from the tensor algebra 
$T(F)$ to $\oplus_n {\cal F}(n)$ such that the Laurent expansions
around the partial diagonals are given by the OPE and such that the
order of the poles at 0 and $\infty$ obeys the restrictions mentioned above. 
Such a map will be interpreted as a system of generalized $n$-point functions
$\langle v_1|\phi^1(z_1)\ldots\phi^n(z_n)|v_2\rangle$. The R states on the
OPE are defined analogously. Of course the distinction between
NS and R representations only makes sense for $F_f\neq 0$. In this case,
the basic representation on $F$ itself is a NS representation.
 
The Laurent expansions around $z=0$ and $z=\infty$  of the $(n+1)$-point 
functions $\langle v_1|\phi(z)\phi^1(z_1)\ldots\phi^n(z_n)|v_2\rangle$
yield states $v_1\otimes (\phi_m v_2)$ and 
$(v_1\phi_m)\otimes v_2$ on the OPE. Thus every state generates a vector
space of states with right and left actions of the Fourier components of
the holomorphic fields. This vector space of states will be called a
representation of the OPE. The space can be graded by the constants $c_1,c_2$
introduced above. For rational theories, its homogeneous subspaces have
finite dimensions. The definition of direct sums and of
irreducible representations is the standard one. For irreducible
representations, the vector space of states can be written as 
$H_i^\ast\otimes H_i$, and as usual we identify $H_i$ with the representation
space. When one works harder, one can define tensor products of 
representations, but we will not need them.

In general a holomorphic OPE has various irreducible representations. A simple
example is given by the basic representation of any theory with 
$F_f\not= 0$. By the field-state identity $H\simeq F_b\oplus F_f$.
When restrict $F$ to $F_b$, the representation on $F_b\oplus F_f$ decomposes
into irreducible representations on $F_b$ and on $F_f$. The first one
is the new basic representation, since $I\in F_b$. The representations cannot
be isomorphic, since $F_b$ is graded by the integers, whereas the eigenvalues
of $L_0$ in $F_f$ are half-integral.

For a vector $v$ of lowest degree $h$ in any representation one has
$L_mv=0$ for $m<0$, since there are no vectors of degree $m+h$. 
With $L_0v=hv$ one finds for normalised $v$
$$
\langle v|T(z)|v\rangle = hz^{-2} 
$$
and
$$
\langle v|T(z)T(w)|v\rangle = {c\over 2}(z-w)^{-4}
+2h(zw)^{-1}(z-w)^{-2}+h^2(zw)^{-2}.
$$
The calculation of $\langle v|U(z)|v\rangle$ from this formula and eq. 
\req{U} is easy and left to the reader. Using $U=0$ in the $(2,5)$ minimal 
model yields $h(h+1/5)=0$. Indeed one can construct two
representations of the OPE of this model, the basic representation
with $h=0$ and another one with $h=-1/5$.
Irreducible representations of minimal models are determined up to isomorphism
by the lowest eigenvalue of $L_0$, but this will not be shown here.

The free fermion theory has a unique irreducible Ramond representation.
Recall that $\sqrt{zw}\langle v_1|\psi(z)\psi(w)|v_2\rangle$ is meromorphic,
with a single pole at $z=w$ of residue $\langle v_1\mid v_2\rangle$.
Taking into account the antisymmetry under exchange of $z,w$ this
yields for a normalised Ramond groundstates $v=v_1,v_2$ 

$$\sqrt{zw} \langle v |\psi(z)\psi(w)| v\rangle =
\frac{z+w}{2(z-w)}.$$

For the Virasoro field $T$ one obtains
$$\langle v|T(z)|v\rangle = (4z)^{-2}$$
thus 
\beq{16}
L_0v = \frac{1}{16}v.
\eeq
  
The bosonic part $F_b$ of the free fermion theory yields the $(3,4)$ minimal
model with $c=1/2$. In the NS case, $H$ decomposes into a direct sum of   
the basic representation $F_b$ with $h=0$ and $F_f$ with $h=1/2$.
We just have seen that the R case yields a representation
with $h=1/16$. Using similar arguments as for $c=-22/5$ one can show
that these are the only irreducible representations.

When one has $N$ free fermions, the $R$ ground states have $h=N/16$ and
form the spin representation of the $so(N)$ Lie algebra obtained above.
For even $N$ the dimension of this space is $2^N$ and the $\Z_2$-decomposition
$H_R=H^0_R\oplus H^1_R$ yields half-spinor representations on the ground states
of $H^0_R$ and $H^1_R$. Similaly, one has the $\Z_2$-decomposition
$H_{NS}=H_{NS}^0\oplus H_{NS}^1$ given by $F^N=F^N_b\oplus F^N_f$.
For $N=16$, the fields corresponding to the $R$ groundstates have $h=1$,
and one can find a new holomorphic theory with $F\simeq H_{NS}^0\oplus H^0_R$.
One has 
$$\dim\,F(1)= 120+2^7,$$
so it is not hard to guess that this is
the theory of $E_8$ currents with OPE \req{currentOPE} and 
$d_{ab}=\delta_{ab}$. The Virasoro field of this theory lies in $H_{NS}^0$
and restricts to the Virasoro field of the $so(16)$ theory, with $c=8$.

When one takes two copies of this theory, the sums of the currents
generate a subtheory, and by eq. \req{ctilde} the complementary
theory has $\tilde c(E_8,2)=1/2$. Indeed it is isomorphic to the $(3,4)$
minimal model. Thus there is a close link between this minimal model and $E_8$. 
This should make eq. \req{e8} somewhat less miraculous. 
 
An important property of conformally invariant quantum field
theories is the possibility to transfer them to arbitrary compact
Riemann surfaces. This is possible since operator product expansions
are local and can be rewritten in terms of conformally equivalent
local coordinates. In particulars one can find a new 
Virasoro field $\tilde T$ adapted to the changed coordinates.
Let us put $z=f(x)$, $w=f(y)$ and evaluate $\tilde T$
with respect to $x$. One has

\begin{eqnarray*}
(z-w)^2&=&f'(x)f'(y)(x-y)^2\\
&&-{1\over 24}\left(2f'(x)f'''(x)-3f''(x)^2+2f'(y)f'''(y)-3f''(y)^2\right)
(x-y)^4\\ &&+\,O\left((x-y)^6\right).
\end{eqnarray*}

By insertion in the OPE of $T$ , one easily confirms that
$$\tilde T(x)= T(z)(dz/dx)^2+{c\over 12}S(f)$$
where $S(f)=(f'f'''-3(f'')^2/2)(f')^{-2}$ is the Schwarzian derivative.
For $z=\exp(ix)$ we have $S(f)=1/2$, which shows that the holomorphic
part of the generator of time translations on $S^1\times \R$ is given by 
$$-\int \tilde T(x)\,\frac{dx}{2\pi}=L_0-c/24.$$

By the OPE, $n$-point functions on arbitrary compact Riemann surfaces can be 
calculated in terms of the 1-point functions. The one point functions on a  
torus with periods $(2\pi, 2\pi\tau)$ and $q=\exp(2\pi i\tau)$ have the form 
$$\langle\phi\rangle_\tau=
Tr \left(q^{L_0-c/24}\,\bar q^{\bar L_0-\bar c/24}\,\phi(x)\right),$$
and are independent of $x$. 

For $\phi=I$ one obtains the partition function 
$Z(\tau)=\langle I\rangle_\tau$. By conformal invariance, $Z$ only depends on 
the ratio $\tau$ of the torus periods. By choosing a different set of 
generators of the torus periods, this ratio can be changed to
$(A\tau+B)/(C\tau+D)$, where ${AB\choose CD}\in SL(2,\Z)$. This explains
the modular invariance of the partition functions of bosonic CFTs. In the
fermionic case the spin structures of the torus have to be taken into
account. We do not need to discuss this case, since one always can restrict
a fermionic CFT to its bosonic part.
 
For rational CFTs there are finite sets of irreducible representations for 
the OPEs of the holomorphic and the anti-holomorphic fields. Let the
corresponding representation spaces be labelled by $V_i$, $\bar V_j$,
where $V_0$, $\bar V_0$ are the basic representations to which the vacuum
belongs. Let $V_i\otimes\bar V_j$ occur with multiplicity $n_{ij}$ in $H$.
Then 
$$Z=\sum_{i,j} n_{ij} \chi_i\bar\chi_j,$$
where
$$\chi_i(\tau)=Tr_{V_i}\ q^{L_0-c/24}$$
and analogously for $\bar\chi_j$. This explains eqs. \req{chi}, \req{Z}.

For the free fermion OPE, we had found a NS representation on a space $V_{NS}$
and a R representation on a space $H_0\otimes V_R$ where $V_{NS}, V_R$ are
spanned by vectors of the form $e_{k_1}\wedge e_{k_2}\wedge\cdots e_{k_n}$
with $k_1>k_2>\cdots k_n>0$ and $e_k=\exp(-ikx)$, with half-integral and
integral $k$, resp. The $\Z_2$ degree of such a vector is 0 for even $n$ and
1 for odd $n$. The two-dimensional space $H_0$ has degree 0 and 
one-dimensional even and odd subspaces. This yields characters 
\begin{eqnarray*}
\chi^{(3,4)}_{NS}&= & q^{-1/48}\prod_{r=1/2,3/2,\ldots} (1+q^r)\\
\chi^{(3,4)}_R&= & 2q^{1/24}\prod_{n=1}^\infty (1+q^n),
\end{eqnarray*}
where we used $c=1/2$ and in the Ramond case the fact that $L_0$ has eigenvalue 
$1/16$ on the ground state. The factor of 2 in $\chi^{(3,4)}_R$ comes from
the $\Z_2$ grading.

When we restrict the OPE algebra of the holomorphic free fermion theory to 
its bosonic part, one obtains the OPE of the $(3,4)$ minimal model. The
decomposition of the free fermion characters into irreducible $(3,4)$
characters is given by

\begin{eqnarray*}
\chi^{(3,4)}_{NS}&=&\chi^{(3,4)}_0+\chi^{(3,4)}_2\\
\chi^{(3,4)}_R&= & 2\chi^{(3,4)}_1,
\end{eqnarray*}
where
$$\chi^{(3,4)}_0-\chi^{(3,4)}_2= q^{-1/48}\prod_{r=1/2,3/2,\ldots} (1-q^r).$$
Up to phases, the $\chi^{(3,4)}_i$, $i=0,1,2$,
are invariant under the subgroup $\Gamma_0(2)$ of the modular group which is
given by matrices ${AB\choose CD}\in SL(2,\Z)$ with even $C$.
When one combines holomorphic and anti-holomorphic parts of the full  minimal 
$(3,4)$ CFT one obtains a torus partition function 

\beq{z34}
Z^{3,4}=|\chi^{(3,4)}_0|^2+|\chi^{(3,4)}_1|^2+|\chi^{(3,4)}_2|^2
\eeq
which is invariant under the full modular group. 

Apart from their product representations, $\chi^{(3,4)}_{NS}$ and
$\chi^{(3,4)}_R$ also have sum representations, which are more important
for our present purpose. The spaces $V_{NS}$ and $V_R$ are direct sums
of subspaces $V_{NS}(n)$ and $V_R(n)$ spanned by vectors of the form
$e_{k_1}\wedge e_{k_2}\wedge\cdots e_{k_n}$ with fixed $n$. We can 
put  $k_m=k_{m+1} +1+s_m$, where $k_{n+1}=b-1/2$ and $b=0$ in the NS and 
$b=1/2$ in the R case. Then

$$\sum_{m=1}^n k_m = n^2/2 + bn+ \sum_{m=1}^n ms_m.$$
The $s_m$ are independent and take values from 0 to $\infty$. 
Thus 
$$Tr_{V_R(n)}\ q^{L_0-1/16} = \frac{q^{n(n+1)/2}}{(q)_n}$$
and

$$Tr_{V_{NS}(n)}\ q^{L_0} = \frac{q^{n^2/2}}{(q)_n}.$$
Here $(q)_n = (1-q)(1-q^2)\cdots(1-q^n)$ is the
so-called $q$-deformed factorial. For small $\tau$ it behaves like
$(-2\pi i\tau)^n n!$, which explains the name.

For the (2,5) minimal model, one also finds characters with nice 
representations as sums and products. The space of all holomorphic fields is
spanned by expressions of the form 

$$\partial^{m_n}T\ldots\partial^{m_2}T\,\partial^{m_1}T$$
where $m_n\geq\ldots \geq m_1$ and the normal ordering is suppressed. By eq.
\req{U} and $U=0$ we have
$$T(z)T(w)=-{3\over 10}\partial^2 T(w) + O(z-w).$$
Taking derivatives $(\partial_z+\partial_w)^{2n}$ and
$(\partial_z+\partial_w)^{2n-1}$ of this equation, one sees that the normal
ordered products $N_0(\partial^nT,\partial^nT)$ and 
$N_0(\partial^nT,\partial^{n-1}T)$ are linear combinations of terms which are
simpler in a suitable lexikographical order. One can show that no further 
linear dependencies exist. Thus the character in the vacuum sector is 
given by the sum over all expressions of the form
$$q^{2+m_n}\ldots q^{2+m_2}q^{2+m_1}$$
where $m_k\geq m_{k-1}+1$ and $m_1\geq 0$. Together with
the factor $q^{-c/24}$ this yields
$$\chi^{(2,5)}_0= q^{11/60} \sum_{n\in\N} \frac{q^{n(n+1)}}{(q)_n}.$$

In the sector with ground state $v$ and $L_0v=-v/5$ one
finds an anlogous formula. Here $L_1v$ does not vanish,
but again terms with $L_nL_n$ and $L_nL_{n-1}$ are linear
combinations of simpler terms. Thus the character in this
sector is given by the sum over all expressions of the form

$$ q^{1+m_n}\ldots q^{1+m_1}q^{1+m_1},$$
where $m_k\geq m_{k-1}+1$ and $m_1\geq 0$. Together with
the factor $q^{-c/24-1/5}$ this yields

$$ \chi^{(2,5)}_1= q^{-1/60} \sum_{n\in\N} \frac{q^{n^2}}{(q)_n}.$$
Since the work of Rogers and Ramanujan it is well known that $\chi^{(2,5)}_0$
and $\chi^{(2,5)}_1$ are modular and have the product expansions \req{RR}.
The partition function
\beq{Z25}
Z^{2,5}=|\chi_0^{(2,5)}|^2+|\chi_1^{(2,5)}|^2
\eeq
is invariant under all modular transformations.

The sum forms of the (2,5) characters look very similar to the 
the free fermion ones. Can they be generalized? In both cases one has
expressions 

$$\sum_{n\in\N} \frac{q^{an^2/2 +bn+h}}{(q)_n},$$
where $a$ characterizes the CFT, and $b,h$ change according to the sector of
the theory. When the tensor product of $r$ models is formed, one obtains 
the product of the corresponding characters, thus

$$\chi = \sum_{n\in\N^r} q^{nAn/2 +bn+h}/(q)_n,$$
where now $A$ is a diagonal $r\times r$ matrix, $b\in \Q^r$ and
$bn$ is to be understood as scalar product in $\Q^r$. Finally

$$(q)_{n_1,\ldots,n_r} = \prod_{i=1}^r (q)_{n_i}.$$
As has been mentioned in the introduction, one can find other characters of 
conformally invariant quantum field of the form \req{chiQ}, but with a
non-diagonal matrix $A$. For the $(2,q)$ minimal models one obtains the
Andrews-Gordon generalization of the Rogers-Ramanujan identities, and for
$A=2{\cal C}(E_8)^{-1}$ and ${\cal C}(E_8)$ the Cartan matrix of $E_8$
one obtains \req{e8} as an equation for $\chi^{(3,4)}_0$.

To see how the matrices $A$ arise from a given conformal field theory,
we have to look at its massive perturbations. This is done in the
next section. It needs more physics and leads somewhat beyond the domain of
present-day rigorous mathematics. A reader who prefers to accept equation
\req{chiQ} without physical motivation can skip the next section.

\section{Integrable Perturbations}

To explain the form \req{chiQ} of characters, the corresponding CFTs
have to be understood as limits of more general integrable quantum field
theories. The latter are still local, but no longer conformally invariant. 
They are not under complete mathematical control, though integrability helps 
a lot. Their partition functions and some properties of their $n$-point 
functions are calculable, but the calculations depend on assumptions which 
are plausible but not proven.

We shall proceed as follows. First we shall discuss the theory of free
massive fermions as a perturbation of a conformally invariant theory
of massless fermions.
Since the theory is free, everything can be calculated exactly with
little effort. We shall see that the (3,4) minimal models has perturbations
with two parameters, one given by the fermion mass just mentioned, the other
one related to $E_8$. If only the latter one is used, the theory remains
integrable, due to the existence of suitable higher conservation laws.
Then we sketch the description of the state space of massive
theories, the scattering matrices of integrable theories and the Bethe
ansatz. The form of the characters \req{chiQ} will follow and we shall see
how the integrable perturbation determines the matrix $A$, at least in
principle. Much more could be done, since the mathematical structure 
of the integrable theories is very beautiful. Its relations to algebraic
K-theory have not been explored yet, however.

Nevertheless, the physical arguments explained in this section can be used 
as a guide for future mathematical research. Mathematicians who want to follow
this lead will have to learn to read some physics textbooks. Thus this
section is less self-contained than the others and uses some elementary
physics background and terminology, for example from relativistic mechanics. 
 
For free massive fermions the quantization procedure discussed at the
beginning of section 2 can be used. The Dirac equation \req{Dirac}
couples $\psi_L$ and $\psi_R$, so both have to be considered together.
Let us first do that in the massless case. We identify the holomorphic field 
$\psi$ used above with $\psi_L$. Recall that we use a splitting
$V = V_+\oplus V_-$ of the vector space to be quantized. The space $V_-$ 
corresponds to the negative Fourier components with respect to time
translation, and a correct description of time demands that this is done
consistently for all fields. Now for right- and left-movers a given time 
translation corresponds to opposite space translations. 
We described the distribution $\langle \psi_L(x)\psi_L(y)\rangle$
as the $\epsilon\rightarrow +0$ limit of a function
$\langle \psi_L(x)\psi_L(y+i\epsilon)\rangle$, with holomorphic dependence
on $y+i\epsilon$. Thus $\langle \psi_R(x)\psi_R(y)\rangle$ is the limit
of a function $\langle \psi_R(x)\psi_R(y+i\epsilon)\rangle$ which depends
holomorphically on $(-y+i\epsilon)$ or anti-holomorphically on
$y+i\epsilon$. The latter description allows us to obtain all distributions
$\langle \phi^1(x_1,t_1)\ldots\phi^n(x_n,t_n)\rangle$ by analytic continuation 
of a euclidean $n$-point function with purely imaginary time components $t_k$. 
The limit is taken such that $\Im t_1<\ldots<\Im t_n$.

For the $n$-point functions we still can use Wick's theorem \req{Wick},
with $\langle \psi_L\psi_R\rangle=0$ and
$$\langle\psi_R(z)\psi_R(w)\rangle =(\bar z -\bar w)^{-1}.$$
Since $\psi_R$ is anti-holomorphic in the euclidean description, one
usually uses the notation $\psi_R=\bar\psi$.  We will follow this notation, 
though it is somewhat misleading, since $\psi_L$ and $\psi_R$ are independent
fields which are both real with respect to the standard anti-involution of $F$.

Now we consider deformations of a given quantum field theory. We assume that
the deformed theories are translationally invariant and have a unique vacuum
vector $1\in H$. We assume that we have euclidean $n$-point functions which are
real analytic, apart from singularities along the partial diagonals. From the
latter we read off the operator product expansion. It has the form

$$\phi_i(z)\phi_j(w)=\sum_k f_{ijk}(z-w)\phi_k(w),$$
with real analytic functions $f_{ijk}(z-w)$. We assume rotational
invariance, such that every field $\phi_i$ has a conformal spin
$s_i$ analogous to $h_i-\bar h_i$. This means that

$$(z-w)^{s_i+s_j-s_k}f_{ijk}(z-w)$$
only depends on the radial distance $|z-w|$. The dependence on this distance
is far more complicated than in the conformally invariant case, however. In
particular, the space $F$ of local fields is no longer graded by a conformal
dimension $d=h+\bar h$. Nevertheless we assume that at short
distance the breaking of conformal invariance is weak. This means that
$F$ is filtered by a scaling dimension $d$, such that

$$|f_{ijk}(z-w)| = o\left(|z-w|^{d_k-d_i-d_j+\epsilon}\right)$$
for all $\epsilon>0$.
 
In this way one gets close to an axiomatic definition of general
quantum field theories, but one is very far from calculability.
Efficient calculations are possible, when a quantum field theory
depends differentiably on some parameter $\lambda$, such that for
$\lambda=0$ all $n$-point functions are calculable. The deformation  
away from $\lambda=0$ is performed by perturbation theory. In our
case, the unperturbed theory will be conformally invariant.

In general, the structure of quantum field theories is so restricted that the
perturbations of a given theory can be described by points in a low
dimensional moduli space. We consider a single parameter $\lambda$ for ease
of notation. The space of fields $F$ should be locally trivializable over
the parameter space such that the $n$-point functions of the deformed theory
depend real analytically on $\lambda$. The $n$-point functions are constrained
by the requirement that expansions around the diagonals of $\C^n$ does not
lead to additional fields. 

We have an  OPE
$$\phi_i(z)\phi_j(w)=\sum_k f_{ijk}(z-w,\lambda)\phi_k(w).$$
At the conformally invariant point $\lambda=0$, eq. \req{scaling} yields

$$f_{ijk}(z-w,0)=C_{ijk} (z-w)^{h_k-h_i-h_j}
(\bar z- \bar w)^{\bar h_k-\bar h_i-\bar h_j},$$
with constant coefficients $C_{ijk}$.
In a first order expansion around $\lambda=0$ this yields equations 
\begin{eqnarray}\la{def}
\partial_\lambda\langle \phi_i(z)\phi_j(w)\ldots\rangle &=& \sum_k
C_{ijk}(z-w)^{h_k-h_i-h_j}(\bar z-\bar w)^{\bar h_k-\bar h_i-\bar h_j}
\partial_\lambda\langle\phi_k(w)\ldots\rangle\nonumber\\&&
+\sum_k\, \langle\phi_k(w)\ldots\rangle \partial_\lambda f_{ijk}(z-w,0) .
\end{eqnarray}
These are consistency equations for the $\lambda$-derivatives of the
$n$-point functions and of the OPE functions $f_{ijk}$ at $\lambda=0$.
By eq. \req{scaling}, the coefficients of these $\lambda$-derivatives
are homogeneous under scaling, thus one also has a basis of solutions of the
consistency equations which is homogeneous under scaling. In other words, 
we can assume that $\lambda$ itself has
well defined scaling dimension $\delta,\bar \delta$. Since the conformal 
spins $h(\phi)-\bar h(\phi)$ are integral and cannot change, one needs
$\delta=\bar \delta$. We assume $\delta\geq 0$, for a reason which will be
explained below.

Now we consider the free massless fermion with components $\psi,\bar\psi$ and
show that the massive Dirac equation \req{Dirac} is the only possible
deformation. Note that we demand that the deformed theory still has fields
$\psi,\bar\psi$ which generate all of $F$ by normal ordered products.
For non-zero $\lambda$ the local field $\bar\partial\psi$ no longer will 
vanish, but a priori one will have
$$\bar\partial\psi = \lambda\chi^1 + \lambda^2\chi^2 +\ldots$$
Comparing scaling dimensions $h$ one finds $1/2=k\delta+h(\chi^k)$,
$1=k\delta+\bar h(\chi^k)$. The only possible solution is $\delta=1/2$,
$$\bar\partial\psi = \lambda\bar\psi,$$
and analogously for $\partial\bar\psi$, up to a constant. Note that
higher powers of $\lambda$ cannot occur, since there are no fields
$\chi^k$ with appropriate dimensions. With a suitable
rescaling of the fields the constant can be chosen to be 1 or $-1$.
Since euclidean $n$-point functions have to go to zero at infinity only
the plus sign in $\partial\bar\partial\psi=\pm\lambda^2\psi$ is allowed,
such that 
$$\bar\partial\psi = \lambda\bar\psi.$$

In this way we recover the theory of a free fermion on $S^1\times \R$ which
satisfes the Dirac equation \req{Dirac} with non-vanishing $\mu=\lambda$. 
It can be quantized by the same method as the massless case. 
The theory is manifestly invariant under space and time translations, thus
the 2-point function of the fermion field depends on two variables $x,y$. 
It is easily obtainable from  the Green's function of $\partial^2_x
+\partial^2_y-\mu^2$. On $\R^2$ this is a 
Bessel function depending on $\mu |z-w|$. More details can be found in any 
textbook of quantum field theory and will not be given here.  Note, however,
that the short distance behaviour $z-w\rightarrow 0$ is equivalent to the
massless limit $\mu\rightarrow 0$. If one considers the theory on 
$S^1\times \R$ instead of $\R^2$, the leading short distance singularities do 
not changer. This can be understood as a consequence of the locality of the 
theory.

For general deformations with $\delta\neq 0$, the two-point function
will depend on $\lambda|z-w|^{2\delta}$. For negative $\delta$,
the short distance behaviour would be more singular than for the unperturbed
theory, and an infinite Taylor expansion in $\delta$ would introduce
short distance singularities which are stronger than any negative power of
$|z-w|$. Presumably this is inconsistent, and in any case it could not be
handled by available methods, so we exclude $\delta<0$. The case $\delta=0$ 
includes the deformations within the moduli spaces of the conformally
invariant theories themselves. It is particularly important, but not in our
context and will not be considered here.

We now need to answer two important questions: How does one
classify perturbations with $\delta>0$ and which of them are integrable?
Both questions turn out to be related to the study of derivatives.
When a derivative $\partial\phi$ is non-zero, there is nothing much to
discuss, one just can choose the trivialization of $F$ over the moduli
space of the perturbed theory such that relations like $\partial\phi=\chi$ 
between $\phi,\chi\in F$ remain true when $\lambda$ is varied.

Holomorphic fields $\phi$ need not remain holomorphic, however, as we have 
seen. Since dimensions are bounded from below, $\bar\partial\phi$ is a 
polynomial in $\lambda$. Let us assume that this polynomial is linear, as it 
was the case for the free fermion fields. In particular this is enforced if
the theory is unitary and $\delta>1/2$. Let $F(h,\bar h)$ be the space of 
fields with conformal dimensions $h,\bar h$. We define a map 
$$\vartheta: \ F(h,0)\rightarrow F(h-\delta, 1-\delta)$$
by $$\bar\partial\phi = \lambda\,\vartheta\phi.$$

A particularly important case concerns the energy momentum tensor $T,\bar T$.
Energy and momentum must be conserved to insure translation invariance
in time and space. In the coordinates $x,t$ conservation laws have the
form $\partial_t\phi=\partial_x\chi$, since this implies that
$\int \phi dx$ is independent of $t$. Thus one expects

\begin{eqnarray*}
\bar\partial T &=& \lambda \partial\Phi\\
\partial\bar T &=& \lambda\bar\partial \bar\Phi.
\end{eqnarray*}
Scaling yields $h(\Phi)=\bar h(\Phi)=1-\delta$ and $h(\Phi)=h(\bar\Phi)$.
The change of momentum is given by $\int(\Phi-\bar\Phi)dx$. Since the
momentum on $S^1$ is quantized, this has to vanish, so $\Phi-\bar\Phi$
can be written as a derivative with respect to $x$. For dimensional reasons,
this cannot be realized in a non-trivial way, so we need $\Phi=\bar\Phi$.
The field $\Phi$ uniquely characterizes the deformation and each real field 
$\Phi$  with $h(\Phi)=\bar h(\Phi)<1$ generates a possible perturbation
with one parameter $\lambda$. Thus we have classified the massive deformations
of a CFT. For the perturbation of the free fermion theory considered above
one finds $\Phi=N_0(\bar\psi\psi)/2$, as we shall see.

Integrable theories are characterized by more conserved quantities, so we are
interested in more holomophic fields which behave like $T$. In other words,
we want to find fields for which the image of the map $\vartheta$ lies in
$\partial F(h-1-\delta, 1-\delta)$. In particular, this is true for the
fields of the form $\partial\phi$ with $\phi\in F(h-1,0)$, but these fields 
yield no conserved quantities, since $\int\partial_x\phi\, dx=0$.
Conversely, every field with vanishing integral over $S^1$ is of this form.

Put $D(h,\bar h)=\dim F(h,\bar h)$ and
$$\Delta(h)=\left(D(h,0)-D(h-1,0)\right)-
\left(D(h-\delta, 1-\delta)-D(h-1-\delta, 1-\delta)\right).$$ 
When $\Delta(h)>0$, the subspace of $F(h,0)/\partial F(h-1,0)$ for which the 
image of $\vartheta$ lies in 
$\partial F(h-1-\delta, 1-\delta)/ \partial \vartheta F(h-1,0)$ has at least  
dimension $\Delta(h)$. We have seen that this subspace  yields conserved 
quantities in the perturbed theory. This is Zamolodchikov's counting
argument.

In the $(3,4)$ minimal model with partition function \req{z34} there are
two fields with $h=\bar h<1$, namely $\psi\bar\psi$ and a field with conformal 
dimensions $(1/16,1/16)$. The latter can be used for a perturbation with
$\delta=15/16$. The $D(h,\bar h)$ are known from eq. \req{z34}
and one finds conserved quantities corresponding to holomorphic fields
of dimensions $h_i=2,8,12,14,18,20$. For $h_i=2$ this is just the 
energy-momentum, but the higher conserved quantities yield an integrable 
theory. The relation to the Coxeter exponents
$1,7,11,13,17,19,23,29$ for $E_8$ is obvious, and there are strong
arguments for the existence of conservation laws for fields of conformal
spin $m_i+30n+1$, where $m_i$ is an $E_8$ Coxeter exponent and $n\in \N$
\cite{Z89, Z91}. Note that 30 is the Coxeter number of $E_8$. 

The conservation laws of the theory of free massive fermions follow the same
pattern. In this case Zamolodchikov's counting argument does not apply. Since
$\mu$ has dimensions $(1/2,1/2)$ and the perturbation of a conservation law
$\bar\partial\phi=0$ yields
$$\bar\partial\phi=\mu\chi^1+\mu^2\chi^2,$$
with $h(\chi^2)=h(\phi)-1$ and $\bar h(\chi^2)=0$. Thus it is not sufficient 
that $\chi^1$ is a derivative field. Instead, one can use the fact that the 
theory is free to find explicit conservation laws. One writes
the relevant cases of the OPE in the form
$$\partial^m\psi(z)\partial^n\psi(w)=
\langle\partial^m\psi(z)\partial^n\psi(w)\rangle\,I +
:\partial^m\psi\partial^n\psi:(w)+o(|z-w|^0),$$
and analgously for $\bar\psi$. In the limit $\mu=0$, the normal ordering
by $::$ coincides with $N_0$. Since $\partial^n\psi\in F_f$ one has
$:\partial^n\psi\partial^n\psi:=0$ and consequently for $n\geq 1$
$$\bar\partial:\partial^n\psi\partial^{n+1}\psi:=
\mu^2 :\partial^{n-1}\psi\partial^{n+1}\psi:=
\mu^2 \partial:\partial^{n-1}\psi\partial^n\psi:\,.$$
For $n=0$ one finds
$$\bar\partial:\psi\partial\psi:=\mu\partial:\bar\psi\psi:\,,$$
that is $\bar\partial T=\mu\Phi$, where 
$$\Phi=\frac{1}{2}:\bar\psi\psi:\,.$$

Thus we have conserved quantities for fields of conformal spin $m+2n+1$,
where $n\in \N$, $m=1$ is the unique Coxeter exponent of $A_1$ and 2 is the
Coxeter number of $A_1$. Thus the pattern of conserved quantities is
analogous to the $E_8$ perturbation. For arbitray ADE Lie algebras $X$ one
expects such a pattern for a perturbation of the theory with field space
$\tilde F^2_X$ discussed in the previous section. 

We now could go on to the description of the Bethe ansatz for calculations
in integrable massive theories, but let us digress briefly to see how
perturbed $n$-point functions are calculated. The main purpose of the
digression is to convince mathematicians that quantum field theory may
be difficult, but is certainly no black art. The eqs. \req{def} form a
huge system, but conjecturally all solutions are known. There is a big space
of trivial solutions, which just comes from the possibility to relabel the
fields by acting with some $\lambda$-dependent elements of $GL(F)$. 
Surprisingly, this will turn out to be important, but of course we are only
interested in solutions modulo the trivial ones. The interesting solutions
are the ones obtained from the fields $\Phi$ considered above. Not much
details will be given, but the reader can check the result for the free 
massive fermion, 
(by Wick's theorem, the $n$-point functions of $\psi,\bar \psi$ are
determinants of two-point functions, so it is sufficent to study the latter).

One would like to put
\beq{prov}
\partial_\lambda \langle\phi_1(x_1)\cdots\phi_n(x_n)\rangle=
\int dx \langle\Phi(x)\phi_1(x_1)\cdots\phi_n(x_n)\rangle,
\eeq
since one needs an expression which is linear in $\Phi$ and preserves
translational invariance and locality. In general, however, the integral on 
the right hand side is divergent at the partial diagonals $x=x_i$,
$i=1,\ldots,n$, and at infinity. We first regularise the divergence at
infinity. One expects that the $x$-integration can be restricted to the
domain $|x|<R$, when one componsates by adding a correction term
$$\langle v(R)|\phi_1(x_1)\cdots\phi_n(x_n)\rangle$$
to the right hand side of eq. \req{prov}. By the OPE expansion, $v(R)$ can
be determined from the special case $n=1$. For 1-point functions we need
\beq{1pert} 
\partial_\lambda\langle\phi(0)\rangle=0
\eeq
for all $\phi\in F$, which determines $v(R)$ uniquely. By eq. \req{infty}, no
divergence at infinity occurs in $\int dx \langle\Phi(x)\phi(0)\rangle$,
when the conformal dimensions of $\phi$ are sufficiently large. Thus we can 
restrict $v(R)$ to a finite dimensional subspace of $H^\ast$, when we take
the limit $R\rightarrow \infty$. When $h(\Phi)+h(\phi)=1$ and 
$h(\phi)=\bar h(\phi)$, one has logarithmic divergences both at 0 and
$\infty$, such that the choice of $v(R)$ must be correlated with the
choice of the regularisation at the diagonal $x=0$, but this is no problem. 

In the following we assume for ease of notation that the $x$-integration
converges at infinity. To handle the other divergences, we exclude 
some $\epsilon$-neighbourhood of the partial diagonals and of infinity from
the domain of integration and denote the integral over the complement of this
domain by $\int_\epsilon$. For finite epsilon, locality is broken, so a limit
$\epsilon\rightarrow 0$ is necessary, but to achieve convergence we need 
renormalisation.

To formulate perturbation theory in the presence of regularisations,
we have to give a differential structure to the space of quantum
field theories with a given vector space of fields $F$.
Let $\C_n$ be the space $\C^n$ minus its partial diagonals.
Let ${\cal F}_n$ be the space of functions $\C_n\times F^n\rightarrow \C$
which are linear in $F^n$, real analytic in $\C_n$ and have sufficiently
good behaviour close to the partial diagonals and at infinity.
A quantum field theory defines an element in ${\cal F}=\oplus_n {\cal F}_n$.
Suppose that the quantum field theory depends on some parameter space $\Pi$
and that the map $\Pi\rightarrow {\cal F}$ is differentiable. Perturbation 
theory is supposed to give the map $T\Pi\rightarrow T{\cal F}$ and its
generalization to higher order jets.

This description is not quite right yet. The elements of $GL(F)$ act on
${\cal F}$ in a natural way. This does not lead to new quantum field theories,
just to a reparametrisation of one and the same theory. Let $T_0{\cal F}$ be
the subspace of $T{\cal F}$ given by the $GL(F)$ action. Then in general
we only can expect that the map $T\Pi\rightarrow T{\cal F}/T_0{\cal F}$ is 
natural.

In the example of massive fermions, $\Pi$ is the positive real axis with
parameter $\mu$, and the one-dimensional vector space $T\Pi$
is generated by $\Phi=:\bar\psi\psi:/2$. More generally, it should be
possible to identify $T\Pi$ with the real fields for which $h=\bar h\leq 1$.
For a field $\Phi$ of this kind, we denote the corresponding element of 
$T{\cal F}$ at $f\in {\cal F}$ by $\partial_{\Phi}f$.

Let $P$ be the projection of $T{\cal F}$ to $T{\cal F}/T_0{\cal F}$. Then

$$P\,\partial_{\Phi}\partial\langle\phi_1(x_1)\cdots\phi_n(x_n)\rangle =
\lim_{\epsilon\rightarrow 0}
\,P\int_\epsilon dx \langle\Phi(x)\phi_1(x_1)\cdots\phi_n(x_n)\rangle$$
yields a natural candidate for a map $TL\rightarrow T{\cal F}/T_0{\cal F}$.
Due to the OPE expansions for $\Phi\phi_k$ the right hand side is well defined.
Indeed, the problems with convergence come from the most singular terms of
the OPE, and the latter can be subtracted without changing the projection.
In other words, one can find $\gamma(\epsilon)\in End(F)$ such that
\begin{eqnarray*}
\lim_{\epsilon\rightarrow 0}\int_\epsilon dx 
&&\Big(\langle\Phi(x)\phi_1(x_1)\cdots\phi_n(x_n)\rangle\\ 
&&-\langle (\gamma(\epsilon)\phi_1)(x_1)\cdots\phi_n(x_n)\rangle-\ldots\\
&&-\langle \phi_1(x_1)\cdots(\gamma(\epsilon)\phi_n)(x_n)\Big)
\end{eqnarray*}
converges.
The renormalisation just introduced is called wave function renormalisation.
It only is sufficient for first order perturbation. For higher orders
one has to take into account that the perturbing field $\Phi$ has to be
renormalised itself. Higher order perturbation is non-unique, since one
can perturb along arbitrary curves in $\Pi$. When we choose
a flat connection on $T\Pi$, we can identify $\Pi$ itself with a subspace
of $F$, at least locally, such that one can perturb along straight lines.
Such a connection is called a renormalisation scheme.
 
End of the digression, we now come to the Bethe ansatz. First we have to 
introduce the scattering matrices of integrable quantum field theories. The
discussion will be very brief, a good pedagogical account is \cite{D98}. 
In our treatment of conformally invariant theories in two space-time
dimensions we first considered systems on a circle with circumference $L$.
Because of scaling invariance, all values of $L$ are equivalent, so we put
$L=2\pi$. When a perturbation introduces a mass parameter $\mu$, the physics
depends on the product $\mu L$. We are particularly interested in the
scale invariant limit $\mu\rightarrow 0$ or equivalently $L\rightarrow 0$.
Nevertheless, we also have to study the opposite limit
$L\rightarrow \infty$, for which the space-time becomes $\R^2$. In this limit
our systems is invariant under Lorentz transformations, which greatly 
simplifies the analysis. The scattering matrix is defined in this situation,
which we will consider now.

The symmetry group of $\R^2$ with metric $dt^2-dx^2$ is the Poincar\'e
group. The massive Dirac equation and the corresponding free field theory
are invariant under this group, and we require invariance for all quantum
field thories on $\R^2$. We have to consider the translations and the group of
Lorentz transformations $SO(1,1)$, which is isomorphic to the additive group 
$\R$. The eigenvalues of the time and space translations are energy and 
momentum,  which we denote by $(\omega,k)$. On irreducible representations of 
the Poincar\'e group, $\omega^2-k^2$ is constant. Because of locality, we do
not want velocities $k/\omega$ which are greater than 1 (the velocity of 
light). Thus we need $\omega^2-k^2=m^2\geq 0$. We introduce a mass gap
$\mu>0$, such that $m\geq \mu$ for all states occuring in the theory, apart
from the vacuum. This is in contrast to CFTs which have
many states with $m=0$. In our case we can parametrize energy-momentum as
$$(\omega,k)=m(\cosh\theta,\sinh\theta).$$
Lorentz transformations act additively on $\theta$, such that the
irreducible Poincar\'e representations are naturally isomorphic to the space
of square integrable functions on the real line with parameter $\theta$.
They are called one-particle spaces and interpreted as state spaces of a 
particle with mass $m$. When one has a conserved field $\phi$ with conformal 
spin $s$ such that $\bar\partial\phi=\partial\chi$, the action of the 
corresponding conserved quantitity on the state space is given by 
$$f(\theta)\rightarrow a\exp((s-1)\theta)f(\theta),$$
where the quantum number $a$ depends on the particle type. For the component 
$T$ of the energy-momentum tensor one has $s=2$ and a conserved quantity 
$\omega+k$, for $\bar T$ the conserved quantity is $\omega-k$. Thus one of the
quantum number arising from conservations laws is the particle mass.

Since matter with non-zero mass cannot move at the velocity of light, and 
velocity now has a purely continuous spectrum, we expect that after a
sufficiently long time any state of finite energy will separate into 
particles which move at different velocities. Thus at large times one can 
count the number of particles
and determine their types. We assume that there is a finite number $r$ of
particle types and index them by a set $I$. For each $i\in I$ we have a mass
$m_i$ and a corresponding one-particle space $H^1_i$, with a natural 
isomorphism to ${\cal L}^2(\R)$ described above.

Let
$$H^1=\oplus_{i\in I}\,H^1_i\,.$$
At large times, particles will be ordered according to their rapidities.
Accordingly, let $T_>(H^1)$ be the subspace of the tensor algebra $T(H^1)$
for which the order of the rapidities corresponds to the order in the
tensor products. It is more conventional to use symmetric and exterior
products, depending on the statistics of the particles.  For interacting 
particles in one space dimension the use of Bose or Fermi statistics is less
natural than for interacting ones, however. One cannot exchange two particles
without moving one through the other, so it is difficult to disentangle
effects of statistics and interaction. Indeed, we will have to consider more
general cases than Bose and Fermi statistics. In our context, statistics does
not matter, since the partial diagonals of velocity space have measure 0 and
$\Lambda H^1_i$ and $S H^1_i$ are isomorphic to $T_>(H^1)$. 

In our description, the behaviour at large negative and large positive times  
yields isomorphisms

$$H\simeq \otimes_{i\in I}\,T_>(H^1).$$
Combining the two isomorphisms yields a unitary transformation
$$S:  \otimes_{i\in I}\,T_>(H^1)\rightarrow
\otimes_{i\in I}\,T_>(H^1),$$
called the scattering matrix. States with 0 and 1 particles are invariant 
under $S$, so $S$ transforms multiparticle states to multiparticle states. 
In general, the scattering of two particles can produce arbitrarily many 
particles. Suppose, however, that one has a conserved field with conformal
spin $s$. On states with momenta $\theta_n$ and particle types $i_n\in\I$ the
eigenvalue of the corresponding conserved quantity is
$$\sum_n a(i_n)\exp((s-1)\theta_n)$$
and does not change by the scattering. Correspondingly, in integrable
theories the number of particles, the rapidities $\theta_n$ and the
quantum numbers $a(i_n)$ do not change in the scattering process. 

In more than one space dimension integrable theories have to be free.
Indeed, consider a situation where two particles pass each other at a large 
distance. Either they do not influence each other at all, in which case the 
theory is free, or one will see a small change in the direction of motion. 
When there is a single space dimension (i.e. two spacetime dimensions), many
non-free integrable models are known.

If two particle types $i,j$ of an integrable theory have identical quantum 
numbers $a(i)=a(j)$, in particular equal masses, scattering processes may
transform one into the other. Theories of this kind are very interesting, 
but will not be considered here. We will assume that the conserved quantum
numbers uniquely specify the particle type.

In any quantum field theory, $n$-particle scattering will approximately
factorize into a product of 2-particle scatterings, when at any given time at
most two particles get close. In integrable theories this factorization is
exact and remains true for all elements of $H$. Indeed, the symmetry operations
given by the higher conserved quantities can be used to translate the particles 
by arbitrary distances. Thus the scattering matrix of an integrable theory
with non-degenerate masses is determined by its restriction to 
$$S_{ij}:\ H^1_i\otimes H^1_j\rightarrow  H^1_j\otimes H^1_i,$$
more precisely to the subspace with $\theta_1>\theta_2$.
When the one-particle spaces are identified with ${\cal L}^2(\R)$, the action
of $S_{ij}$ is diagonal and given by a function $S_{ij}(\theta_{12})$, with
$\theta_{12}=\theta_1-\theta_2$. By unitarity, $|S_{ij}(\theta_{12})|=1$.
Instead of $\theta_{12}$, another natural variable is $\cosh(\theta_{12})$,
since $m_1m_2\cosh(\theta_{12})=\omega_1\omega_2-k_1k_2$.
 
In terms of a Schr\"odinger wave function, a state with two particles of types
$i,j$ with momenta $k_1,k_2$ and rapidities $\theta_1>\theta_2$ is described by
$$\Psi(x_1,x_2)=\exp(ik_1x_1)\exp(ik_2x_2)$$
for $x_1\ll x_2$ and by
$$\Psi(x_1,x_2)=\exp(i\delta_{ij}(\theta_{12})\exp(ik_1x_1)\exp(ik_2x_2)$$
for $x_1\gg x_2$. Exchanging the particle labelling yields
\beq{exchangemod}
\delta_{ij}(\theta)=-\delta_{ji}(-\theta)\quad\mod\quad 2\pi\Z,
\eeq
at least for bosonic particles. For  $\theta_1>\theta_2$ one has
$$S_{ij}(\theta_{12})=\exp\left(i\delta_{ij}(\theta_{12}\right),$$
since the region $x_1\ll x_2$ dominates for large negative and the region 
$x_1\gg x_2$ for large positive times. At finite time there is no singularity
for $\theta_1=\theta_2$, so we expect $\delta_{ij}$ to be real analytic
functions for all $\theta\in\R$. At fixed $\theta$, the value of $\delta_{ij}$
is only determined up to a multiple of $2\pi$, but differences
$\delta_{ij}(\theta)-\delta_{ij}(\theta')$ are uniquely defined real
numbers, since $\delta_{ij}$ is continuous. We can fix $\delta_{ij}$ by
a normalisation at infinite rapidity. Scattering at large rapidity difference
probes the short distance behaviour of the interaction. At short distance,
our $\delta>0$ perturbations become CFTs, for which there is no scattering.
Thus we can put $\delta_{ij}(+\infty)=0$. Then eq. \req{exchangemod} yields
\beq{exchange}
\delta_{ij}(\theta)+\delta_{ji}(-\theta)+2\pi A_{ij}=0,
\eeq
where $A$ is a symmetric $r\times r$ matrix with
$A_{ij}=-\delta_{ij}(-\infty)/(2\pi).$ From the preceding argument one
expects $A_{ij}\in\Z$, but one can incorporate exotic statistics by
admitting non-integral values. 

For a given integrable theory, we can assume the particle spectrum and the
scattering matrix to be known. In particular, the derivatives of the
$\delta_{ij}$ are rational functions of $\cosh(\theta)$. In integrable theories 
which are invariant under space reflections, the scattering matrix elements
have the form
$$S_{ij}(\theta)=\prod_{x\in Q_{ij}}\frac{\sinh((\theta+i\pi x)/2)}
{\sinh((\theta-i\pi x)/2)},$$
with finite set $Q_{ij}\subset\R$. When the theory tends to a rational CFT
at short distance, one even hast $Q_{ij}\subset\Q$. We will see that these
$Q_{ij}$ determine in a direct way the matrix $A$ in \req{chiQ} and thus the 
central object of our study. In principle the converse should be true, too.
Indeed $A$ should characterize a CFT and an integrable perturbation which
reproduces the $Q_{ij}$. It would be nice to find an algorithm which gives the
result in a more direct way.

For integrable theories, systems of particles on a circle of circumference $L$ 
can be described by the Bethe ansatz. One extends the previous description of
the Schr\"odinger wave function to small distances and looks at its phase
change when one follows one particle position around the circle. 
Let us consider particles of types $i(1),i(2),\ldots$ with momenta 
$k_1,k_2,\ldots$. When the particles do not interact, each one can be described
by a wave function $\exp(ik_mx)$. The $k_m$ are quantized in a simple way,
since $kL$ must be an integral multiple of $2\pi$, or a half-integral multiple
for fermions in the NS case. When the phase changes are taken into account
one finds
$$ k_mL + \sum_{n\not= m} \delta_{i(m)i(n)}(\theta_{mn}) = 2\pi N_m,$$
where in the bosonic case the $N_m$ must be integral. 
By eq. \req{exchange} the quantization of the total momentum is not
affected by the interaction, as long as the $A_{ij}$ are integral.

For small $L$, the rapidities of right and left movers are of the order
$\pm\log(2\pi |N_m|/L)$, respectively. Thus for scattering processes
between a left and a right mover the rapidity difference becomes infinite
for $L\rightarrow 0$, whereas for two right- or two left-movers it becomes
independent of $L$. Thus right movers and left movers decouple, except
for statistical effects. For example, one may need a total fermion number
which is even, or the right and left $NS$ and $R$ sectors may be coupled, as 
in the partition function \req{z34}. 

We have seen that in one space dimension, bosons and fermions cannot be
distinguished in the usual way, but a dynamical distinction is possible. For
single $NS$ fermion states, the momentum should be a half-integral multiple
of $2\pi/L$. In this sense, one even can interpolated between bosonic and
fermionic behaviour, by demanding that for a particle of type $i(m)$ one has
$$N_m\equiv b_i\ \mod\ \Z,$$
with arbitrary $b_i$. Values of $b_i$ which are neither integral
nor half-integral imply exotic statistics. If such particles occur, one needs
a balance between left and right to obtain integral total momentum (up to a
factor $2\pi/L$). To obtain a rational CFT in the $L\rightarrow 0$ limit,
one needs of course $b_i\in\Q$. Similar remarks apply to $A_{ij}$. 

It seems that for a given set of $N_m$, the $\theta_i$ are fixed uniquely.
Nevertheless, the Bethe ansatz is incomplete, since the range of the $N_m$ is
not specified. Let us assume that for particles of type $i$ all values
$N\geq b_i$  are allowed. Then we can evaluate the partition function and
its split into holomorphic and antiholomorphic characters in the CFT limit  
$L\rightarrow 0$. In this limit the interaction between left and right movers
should vanish, since the rapidity difference for any pair of right and
left movers is of order $-\log(\mu L)$. Among right movers, energy and 
momentum can be identified,
so after rescaling the energy is given by $\sum_m k_mL/2\pi$, where the
sum only extends over positive $k_m$. Thus the energy shift due to the 
interaction is given by 
$$-\sum_{n\not= m}\delta_{i(m),i(n)}(\theta_{mn})/2\pi =
{1\over 2}\sum_{ij}n_iA_{ij}n_j $$
when there are $n_i$ particles of type $i$. Summing over the possibilities
for the $N_m$ yields the form \req{chiQ} for the corresponding character,
up to a shift $h-c/24$ of the ground state energy.

When we have Fermi statistics for particles of type $i$, such that $N_m, N_k$ 
have to be different when $m\neq k$ but $i=i(m)=i(k)$, a
term $n_i(n_i-1)/2$ is added to the energy, which can be absorved by a
redefinition of $A,b$. 

The matrix $A$ is given by the local interaction behaviour, but $b$ is
a global quantity which can be different in different sectors of the theory.
When $A_{ij}\not\in\Z$ for some $i,j$ or $b_i-A_{ii}/2\not\in\Z$ for some $i$,
one gets a character which does not transform homogeneously under
$\tau\mapsto \tau+2\pi$. Nevertheless, in many cases one finds acceptable
partition functions by averaging sums $\sum_{ij}n_{ij}\chi_i\bar \chi_j$
over the translations $\tau\mapsto \tau+2\pi n$, $n\in\Z$. This averaging
projects out states with exotic values of the total momentum. 
Thus eq. \req{chiQ} has been explained, at least in an intuitive way.

\section{The connection to algebraic K-theory}
 
The preceding discussions indicate that the search 
for modular functions of the form
$$\chi=\sum_{n\in \N^r}\,q^{nAn/2+bn+h-c/24}$$
with a rational symmetric $r\times r$ matrix $A$, $h\in\Q$ and $b\in\Q^r$ will 
be very interesting. To assure convergence, we will assume that $A$ is positive.
 
The requirement that $\chi$ is modular imposes strong restrictions on $A,b,h$.
In this article we wil not consider the restrictions on $b$. In any case,
the CFTs under consideration have a unique $A$, whereas different 
representations of the OPE of $F_{hol}$ yield sectors with different $b,h$.

Recall that $q=\exp(2\pi i/\tau)$ and $\tilde q=\exp(-2\pi i/ \tau)$.
According to eq. \req{tildechi}, a modular character can be written as a sum 
over terms $\tilde q^k$, with real and rational $k$. Let  us check how $\chi$ 
behaves at small $\tau$, where the dominant contribution should come from
$k=-c_{eff}/24$. 

For ease of notation we first consider the case $r=1$, but
generalisation to arbitrary $r$ will be immediate. We have
$$\chi= \oint_C\, \sum_{n\in Z} q^{nAn/2+bn+h}\, x^{-n}
           \sum_{m\in\N^r} \frac{x^m}{(q)_m}\frac{dx}{2\pi ix}$$
where the path $C$ is a small circle around the origin.
The first factor of the integrand can be evaluated by
Poisson summation,

\begin{eqnarray*}
\sum_n\, &&q^{nAn/2+bn+h}\, x^{-n}\\ &&=
\sum_m\int_n\exp\Big(2\pi i\big(\tau(nAn/2+bn+h)-n(u-2\pi im)\big)\Big)dn
\\&&= (i\tau A)^{-1/2}
\sum_m \exp\left(-\frac{(u-2\pi im)A^{-1}(u-2\pi im)}{4\pi i\tau}
+O(\tau^0)\right),\end{eqnarray*}
where $u=\log x$. The integral over $x$ and the sum over $m$ can be combined
into an integral over the simply connected cover of $C$, which yields
an integral over $u$ along a parallel of the imaginary axis.

For the second factor one we have the explicit form
$$\sum_{m\in\N} \frac{x^m}{(q)_m} =
\prod_{n\in \N} (1-xq^n)^{-1}.$$
When $q$ is close to 1, we can approximate the products over $n$
by integrals:
$$\log\prod_{n\in \N} (1-xq^n)^{-1}
\sim -\int_0^\infty dn \log(1-xq^n) 
= -Li_2(x)/2\pi i\tau.$$
The integral over $u$ can be evaluated by the saddle point
approximation. Vanishing of the derivative of

$$\frac{uA^{-1}u}{2}+Li_2(x_i)$$ 
yields
$$A^{-1}u = v$$
where $v=\log(1-x)$. Exponentiation yields
$$x = (1-x)^A.$$
Since $A$ is positive, the right hand side of this equation
equals 1 for $x=0$ and 0 for $x=1$. The left hand side behaves
in the opposite way, such that the equation has a real solution
with $0<x<1$. Moreover, $x$ is algebraic and the corresponding
Rogers dilogarithm $L(x)$ is demanded to be rational.

Altogether we obtain
$$Z\sim \tilde q^{-k},$$
where $k=L(x)/(4\pi^2)$ and
$$L(x) = \frac{uv}{2}+Li_2(x)$$
is the Rogers dilogarithm. Since $k=c_{eff}/24$ we have 
$$c_{eff}=\frac{L(x)}{L(1)}.$$
 
Now $x$ is algebraic and $k$ must be
rational. As explained in Zagier's talk, this happens for three
cases only, namely $A=1$, $A=2$ or $A=1/2$. They correspond to $c_{eff}=
1/2,2/5,3/5$, resp. All of them turn out to be realized in minimal models. 
Recall that $c=1-6(p-q)^2/pq$ for the $(p,q)$ model. The formula for $c_{eff}$
is similar, one has $c_{eff}=1-6/pq$. The models with $q-p=1$ are unitary.
The cases $A=1$ and $A=2$ correspond to the free fermion and the $(2,5)$
minimal model and have been discussed in detail in section 2. The case
$A=1/2$ yields the $(3,5)$ minimal model with $c_{eff}=3/5$.

Now let us consider higher $r$. We have
$$\chi= \oint_C \sum_{n\in Z} q^{nAn/2+bn+h} X^{-n}
\sum_{m\in\N^r}\,\frac {X^m}{(q)_m}\ \prod_{i=1}^r\,\frac{dx_i}{2\pi ix_i},$$
where the path $C$ now is a Cartesian product of small circles around 
the origin. We use the notation $X^m = \prod_i x_i^{m_i}$ and analogously
for $X^{-n}$. As before, Poisson summation of the first factor yields
$$\chi\sim\int \exp\left(-\frac{UA^{-1}U/2+\sum_i Li_2(x_i)}{2\pi i\tau}
\right)\,\prod_{i=1}^r\,du_i\,,$$
where $U=(u_1,\ldots,u_r)$ and $x_i=\exp(u_i)$.

When we apply the saddle point method, vanishing of the derivatives
of $UA^{-1}U/2+\sum_i Li_2(x_i)$ yields
$$A^{-1}U=V,$$
where $V=(v_1,\ldots,v_r)$ and $\exp(v_i)+\exp(u_i)=1$.

The Riemann surface
$$\hat\C=\{(u,v)\in\C\mid \exp(u)+\exp(v)=1\}$$
already appeared in Zagier's talk.
We complete $\hat\C$ at $u=0$ and $v=0$ by adjoining the points $(0,\infty)$
and , which will be useful for bookkeeping. Values of functions at these
points will be assigned when the functions have a unique limit.

When the intergration path for $Z$ is deformed, several saddle points
may appear. They all have to satisfy the equation 
$U=AV$, $(U,V)\in \C^r$. The corresponding contribution to $Z$
is proportional to $\tilde q^{-k}$, where up to an integral summand
$$k = \sum_{i=1}^r L(u_i,v_i)/(4\pi^2).$$ 
Here the function
$$L: S\rightarrow \C/\Z(2)$$
is the analytic continuation of Rogers dilogarithm discussed by Zagier.
It can be characterized by $L(\infty,0)=0$ and
$$dL = (udv-vdu)/2.$$
and has the properties
\begin{eqnarray*}
L(0,\infty)&=&\frac{\pi^2}{6}\\
L(u+2\pi i,v) &=& L(u,v) + \pi iv\\
L(u,v+2\pi iu) &=& L(u,v) - \pi iu.
\end{eqnarray*}
The multivaluedness of $L$ arises from the residues of $d(L+uv/2)=udv$.
Since $dv=du/(\exp(u)-1)$, these residues are multiples of $2\pi i$.
The notation $\Z(2)$ stands for $(2\pi i)^2\Z$ and serves as a reminder the
the proper context should be the theory of motives.

Let $[\hat\C]$ be the free abelian group with basis $[(u,v)]$ for
$(u,v)\in\hat\C$. We extend $L$ to a linear function on $[\hat\C]$. Elements
of $[\hat\C]$ will be denoted by $(U,V)$. The effective central charge
given by the solutions of out equation $U=AV$ is
$$c_{eff} = \frac{6}{\pi^2}\,L(U,V)\quad\mod\quad 24\Z$$
for the dominant saddle point $(U,V)$, whereas the other saddle points yield
$$c-24 h_i= \frac{6}{\pi^2}\,L(U,V)\quad\mod\quad 24\Z$$
for other conformal weights $h_i$ of the CFT. 
F
There is an infinite number of solutions of $U=AV$, and we have to
consider the corresponding values of $L(U,V)$. After exponentiating the
equation we obtain the algebraic equations
$$x_i = \prod_{j=1}^r(1-x_j)^{A_{ij}}.$$
As long as these equations are independent, they only yield a finite
number of solutions. Solutions of $U=AV$ which yield the same $x_i$ are
related by
\begin{eqnarray*}
U' &=&  U +2\pi im\\
V' &=& V +2\pi in,
\end{eqnarray*}
$m,n\in \Z^r$, where $m=An$. We have
$$L(U',V') = L(U,V) + \pi i(mV-nU) + 2\pi^2mn.$$
Since $A$ is symmetric, $mV-nU = nA^tV-nAV = 0$.
We call $A$ even whenever $m=An$, $m,n\in \Z^r$ implies that $mn$ is even.
In this case 
$$L(U',V') = L(U,V)\quad \mod\quad Z(2).$$ 
In general we have $2L(U',V') = 2L(U,V)\quad \mod\quad Z(2)$. 
F
Now we will introduce the extended Bloch group $\hat B(\C)$ as subquotient
of $[\hat\C]$ \cite{N93,N03}. On the latter group we have a natural linear map 
$\sigma: [\hat\C]\rightarrow \C\otimes_{\Z}\C$ induced by
$$\sigma(u,v)= u\otimes_{\Z} v\,-\,v\otimes_{\Z}u,$$
$\sigma(0,\infty)=\sigma(\infty,0)=0$. Let ${\cal P}$ be the kernel of this map.
We define
$$ \hat B(\C)={\cal P}/{\cal P}_0,$$ where the subgroup ${\cal P}_0$ of 
${\cal P}$ is generated by all elements of the forms
\begin{eqnarray*}
(u,v)+(v,u)-(0,\infty)\\
(u-2\pi i,v)+2(u-v-\pi i,-v)+(u,v)\\
\sum_{i=1}^5 (u_i,v_i)-2(0,\infty),
\end{eqnarray*}
where in the last line 
$$u_i=v_{i-1}+v_{i+1}\quad\mbox{for}\quad i=1,\dots 5$$
and $v_0=v_5, v_1=v_6$ for cyclic symmetry.
The first two lines correspond to $[x]+[1-x]=0$ and $2[1-x]+2[1-x^{-1}]=0$
in $B(\C)$, the last to the five-term relation. The elements in ${\cal P}_0$ are
all annihilated by $L$, as one can prove easily by differentiation. In 
particular, the five-term identity asserts 
\beq{5t}
\sum_{i=1}^5 L(u_i,v_i) = \pi^2/3.
\eeq
Thus we can consider $L$ as a map
$L:\hat B(\C)\rightarrow Z/Z(2)$. Note that the obvious involution of ${\cal P}$
which is induced by $(u,v)\mapsto (v,u)$ is not an involution of ${\cal P}_0$.
Instead we have $L(u,v)=L(0,\infty)-L(v,u)$.

The Bloch group is of relevance for us, since due to the symmetry of $A$
all solutions of $U=AV$ yield elements $(U,V)\in {\cal P}$. Since $\hat B(\C)$ is
less well understood that $B(\C)$ we want to relate it to the latter
group. We have seen that there is a map $\hat B(\C)\rightarrow B(\C)$
given by $(U,V)\mapsto \sum_i\,[x_i]$, where $U=(u_1,u_2,\ldots)$ and
$x_i=\exp(u_i)$. We will see that this map is surjective up to
torsion, i.e. the cokernel of the map contains only elements of finite
order. 

Let $\sum_i [x_i]\in B(\C)$ and $y_i=1-x_i$. Then the element
$\sum_i x_i\times y_i\in [\C^\ast\times\C^\ast]$ must lie in the subgroup
generated by elements of the form $[z_m\times z_n]+[z_n\times z_m]$
and $\sum_m r_m[z_m\times z_n]$, where $\prod_m z_m^{r_m}=1$. Thus
$$\sum_i [x_i\times y_i]= \sum_{m,n} C_{mn}[z_m\times z_n],$$
where
$$C_{mn} =R_{mk}F_{kn} + G_{mn},$$
$R,F,G$ are integral matrices, $G$ is even and for all $k$
$$\prod_m z_m^{R_{mk}} = 1.$$
Note that the $x_i,y_i$ must be among the $z_m$. Now we choose logarithms
$w_m$ of the $z_m$, inlcuding $u_i$ for $x_i$ and $v_i$ for $y_i$. Thus
$$\sum_m R_{mk} w_m = 2\pi iq_k$$
for all $k$ and integral $q_k$. By choosing different logarithms, we can
change the vectors $(q_k)$ by integral linear combinations of the $(R_{mk})$,
$m=1,2,\ldots$. We have
$$N\sum_i \sigma[u_i\times v_i]=\sum_{kn}\,Nq_kF_{kn}\sigma[2\pi i\times w_n],$$
for $N\in\N$. After choosing $N$ such that $(Nq_k)$ is a linear combinations of 
the $(R_{mk})$, one can change the choice of the logarithms, in general in 
different ways for the $N$ copies, such that the r.h.s. vanishes. Thus $N[x_i]$
lies in the image of $\hat B(\C)$ in $B(\C)$.

The linear map on $[\hat\C]$ induced by $(u,v)\mapsto (u\bar v - v\bar u)$
factors through the wedge product. Thus ${\cal P}$ is contained in its
kernel. The value $D(\exp(u))$ of the Bloch-Wigner function $D$ discussed 
in Zagier's talk
is the imaginary part of $L(u,v)-(u\bar v - v\bar u)/4$. Thus on
$\hat B(\C)$ the imaginary part of $L$ coincides with $D$, but the real
part yields new information. When $\Im L(U,V)\neq 0$, then $L(U,V)$
and consequently $(U,V)$ have infinite order. For $B(\C)$ there is a converse
to this statement. Since $B(\C)=B(\bar Q)$ we can restrict ourselves to
algebraic numbers. When $\sum_i n_i[x_i]\in B(K)$ for some algebraic number 
field $K$ and $D$ yields 0 for this element and all its Galois conjugates,
then the element is of finite order, see \cite{DZ91}, section 2. Presumably 
this also is true for preimages of such elements in $\hat B(\C)$.

Now we consider our saddle points $(U,V)$ with $U=AV$. Since $A$ is symmetric,
we have $\sum_k u_k\wedge v_k = 0$, such that $(U,V)\in {\cal P}$.
Moreover, the modular character $\chi$  must be a sum rational powers of
$\tilde q = \exp(-2\pi i/\tau)$. In particular this means that 
$L(U,V)/(4\pi^2)$ is rational or equivalently that $L(U,V)$ is an element
of finite order in $\Q/\Z(2)$. This is true for all solutions of $U=AV$, 
including those obtained by Galois conjugation of the solutions of the 
corresponding exponentiated equation. Thus we know that $(U,V)$ maps to 0 in
$B(\C)$, and we expect that it is an element of finite order in $\hat B(\C)$. 
Obviously, the argument is incomplete, but one can certainly expect
to find a real proof along these lines.

Over the real numbers it is known that the any torsion element of the
Bloch group can be represented by a real and rational linear combination
of roots of unity \cite{FS93}. The latter form the field $\Q_{ab}^+$.  
Thus it is a plausible conjecture that for any symmetric matrix $A$
such that $U=AV$ has discrete solutions which are torsion elements
of the Bloch group the corresponding algebraic numbers $\exp(u_k)$
belong to $\Q_{ab}^+$. 

Work on statistical models and related integrable theories has produced a 
partly conjectural list of matrices $A$ belonging to this class which are
related to the Dynkin diagrams of type ADET. These diagrams $A_r,D_r,E_r,T_r$
have $r$ vertices, with $r\geq 4$ for $D_r$ and $r=6,7,8$ for $E_r$.  Recall 
that $r$ is called the rank of $X_r$ and that the corresponding Cartan matrix 
${\cal C}(X_r)$ has off-diagonal entries ${\cal C}(X_r)_{ij}$ which are
equal to -1 when the vertices $i,j$ are linked by an edge of the diagram and
equal to 0 when they are not. Since the ADE diagrams have no loops, one
has ${\cal C}(X_r)_{ii}=2$ for $i=1,\ldots,r$. The Dynkin diagrams of
type $A$ are just rows of vertices with linked neighbours, such that
${\cal C}(X_r)_{ij}=-1$ for $|i-j|=1$ and ${\cal C}(X_r)_{ij}=0$ for $|i-j|>1$.
The  tadpole diagram $T_r$ is obtained by folding $A_{2r}$ diagrams in the 
middle, such that one gets a pairwise identification of the vertices. One has
${\cal C}(T_r)_{rr}=1$, otherwise the matrix elements of ${\cal C}(T_r)$ and
${\cal C}(A_r)$ are the same. Note that ${\cal C}(A_1)=(2)$ and 
${\cal C}(T_1)=(1)$. We need positive definite Cartan matrices, which 
excludes $E_r$ for $r>8$. Indeed, $\det(A_r)=r+1$, $det(D_r)=4$, 
$\det(E_r)=9-r$ and $\det(T_r)=1$. We also need the Coxeter numbers for these
diagrams, which are $h(A_r)=r+1$, $h(D_r)=2r-2$, $h(E_6)=12$, $h(E_7)=18$,
$h(E_8)=30$, $h(T_r)=h(A_{2r})=2r+1$. 

From pairs of ADET Dynkin diagrams $X,Y$ one obtains the matrices 
\beq{cXY}
A(X,Y) = {\cal C}(X)\otimes {\cal C}(Y)^{-1}.
\eeq
When $U=A(X,Y)V$ it is known or conjectured that $(U,V)$ is a torsion
element of $\hat B(\C)$. Let us check this in the simplest examples. For 
rank 1 we have the pairs $(A_1,A_1)$, $(T_1,T_1)$, $(A_1,T_1)$ and $(T_1,A_1)$.
The yield $A=(1),(2),(1/2)$ in agreement with our examples from minimal
models. The general formula for the effective central charge is known or
conjectured to be
$$c_{eff}(X,Y) = \frac{r(X)r(Y)h(X)}{h(X)+h(Y)}.$$ 
This agrees with the previous results for the rank 1 cases. Indeed, for
$A=1$, the equation $x=(1-x)^A$ yields $x=1/2$. Since $[x]+[1-x]=0$
in $B(\C)$, we have $2[1/2]=0$. In $\hat B(\C)$ we can impose $u=v$,
$\exp(u)=1/2$, which yields $(u,u)+(u,u)=(0,\infty)$ and $L(u,u)=\pi^2/12$.
For $A=2$ the equation $x=(1-x)^A$ yields the golden ratio. With $u=2v$
we obtain $5(u,v)=2(0,\infty)$ in $\hat B(\C)$ and $L(u,v)=\pi^2/15$.

For $A=1/2$ we consider more generally $A(X,Y)=A(Y,X)^{-1}$. In special cases
this corresponds to level-rank duality. More generally, replacement of
$A$ by $A^{-1}$ just yields an exchange of $U$ and $V$ in the
solutions of $U=AV$, which is the involution mentioned above.
Note that for an even matrix $A$ its inverse is even, too. Moreover
$L(u,v)+L(v,u)=L(0,\infty)$ yields $L(U,V)+L(V,U)=rank(A)L(0,\infty)$, which
agrees with
$$c_{eff}(X,Y)+c_{eff}(Y,X) = r(X)r(Y).$$

For $X=A_{k-1}$ and $Y$ of $ADE$ type, comparison of eq. \req{cXY} yields with 
eqs. \req{ctilde} and \req{ccheck} yields
\begin{eqnarray*}
c_{eff}(A_{k-1},Y)&=&\tilde c(Y,k)\\
c_{eff}(Y,A_{k-1})=\check c(Y,k).
\end{eqnarray*} 
The theories with field spaces $\tilde F^k_Y$ and $\check F^k_Y$ are
unitary, so this is an equality between two effective central charges. 
For the special case $k=1$, $Y=A_1$ one knows that the theory has characters
of the form \req{chiQ} with $A=A(A_1,A_1)=1$ and for many other cases there
is good numerical or analytical evidence that it is true, too \cite{K87, KN92,
KM93}, so it probably is true in general. This would yield equalities
relating the dilogariths of the corresponding solutions of $U=AV$ with the
conformal dimensions of these theories, and conversely these equalities
would provide good evidence for the conjecture. In the following section
we shall make a first step by finding all solutions of $U=AV$ for
$A=A(A_m,A_n)$ and general $m,n$.

There is a way to prove that $A(X,Y)$ yields torsion elements for any specified 
pair $X,Y$, though it has to be applied separately for each such pair and does 
not give a general proof \cite{Z91, GT95}.  We give a slightly modified
version of the method. First note that $U=AV$ is equivalent to
$$({\cal C}(Y)\otimes I)U = (I\otimes {\cal C}(X))V.$$
Let $C$ be the Cartan matrix of $A^{(1)}_{2(h(X)+h(Y))}$, the Dynkin
diagram of which is a regular polygon. Consider the equation

$$({\cal C}(Y)\otimes I\otimes I - I\otimes I\otimes C)U'
= (I \otimes {\cal C}(X)\otimes I - I\otimes I\otimes C)V',$$
with 
$$U',V'\in\C^{r(Y)}\otimes\C^{r(X)}\otimes\C^P$$ 
and $P=2(h(X)+h(Y))$. With respect to the last factor we write
$U'=(U^1,\ldots,U^P)$ and analogously for $V'$. One easily sees that
Zamolodchikov's equation determines $U^{n+1}$ in terms of $U^{n-1},U^n$.
Zamolodchikov now claims that $U^1,U^2$ can be chosen arbitrarily. In other
words, the pair  $U^{P-1},U^P$ yields $U^1$ again, such that the solution is
periodic in $n$ with period $P$. For the half period $P/2$ one has 

$$U^{n+P/2}_{ij}=U^n_{\sigma(i)\sigma(j)},$$
where $\sigma$ acts as the involutive diagram symmetry on the vertices of
$A_r$, $D_{2r+1}$, $E_6$, and trivially for the other ADET diagrams. This
claim is true for every pair $X,Y$ checked so far, but a general proof is 
lacking. Whenever it is true one obtains continuous families of elements
$(U',V')\in\hat B(\C)$ and no component needs to have an algebraic
exponential. This implies $(U',V')=0$. 
On the other hand we have a special solution 

$$(U',V')=(\underbrace{U,\ldots U}_{P\ times},
\underbrace{V,\ldots,V}_{P\ times}),$$ 
whenever $U=AV$. Thus $(U',V')=P(U,V)$ and $(U,V)$ is an element of finite 
order in $\hat B(\C)$, with an order dividing $P$. When $\sigma$ acts
trivially on $X$ and $Y$, the order is even a factor of $P/2$. In particular,
the products $Pc_{eff}$ and $24P(h(X)+h(Y))h_i$ should be even integers
when $\sigma$ acts trivially, and integral in any case.

\section {Solving the algebraic equations in special cases}

Let us consider the equation $U=AV$ for special cases of $A=A(X,Y)$.
We will be able to find all solutions for $X=A_1$ and also for
$A(A_m,A_n)$, with arbitrary $m,n$. 
Finding the solutions needs methods from Lie algebra theory, but we
shall try first, how far one can get by elementary algebra alone.
This will be sufficient for $X=A_1$, and it will make some of the later
developments easier to understand.

We first rewrite $U=AV$ as ${\cal C}(Y)U={\cal C}(X)V$. In order 
have a chance to find logarithms of algebraic integers we put
$U=-{\cal C}(X)W$. With $Z=\exp(W)$ we obtain
$$Z^{2-{\cal C}(X)} + Z^{2-{\cal C}(Y)} = Z^2.$$
In this form all exponents are positive integers and all coefficients are 1,
so we have a good chane to find algebraic integers.
This equation has many solutions for which some components of $Z$
vanish. Those will be called non-admissible, since they do not yield 
solutions of $U=AV$. When one treats the equations by elementary algebra
it may be useful to find them first, however, since this reduces the
degree of the $Z$ equations. Many of the following considerations
can be regarded as a systematic procedure to eliminate the non-admissible
solutions. 

In general $Z$ has components $z_{ij}$, where $i$ labels the vertices
of the Dynkin diagram of $X$ and $j$ those of $Y$. We first consider
the case $X=A_1$ where the first index is superfluous. Thus for $Y=A_n$
we have variables $Z=(z_1,\ldots,z_n)$. One equation links $z_1,z_2$,
the next ones link $z_1,z_2,z_3$, then $z_2,z_3,z_4$ and so on. To find a
uniform solution we use the semi-infinite $n\rightarrow\infty$ limit of $A_n$ 
and an infinite series of variables $Z=(z_1,z_2,\ldots)$. Generically every 
equation determines $z_n$ in terms of the preceding $z_k$, thus eventually
in terms of $z_1$. To reduce to the case $Y=A_n$ we just have to take the first 
$n$ equations and to put $z_{n+1}=1$. The $D$ and $E$ cases can be handled 
similarly, though the branching needs extra attention. For the resulting
values of $c_{eff}$ see \cite{KM90}.
take the first $n$ equations and to put $z_{n+1}=1$. The $D$ and
$E$ cases can be handled similarly, though the branching needs
extra attention.  

For the $A$ series the equation are
$$1 + z_{i-1}z_{i+1} = z_i^2,$$
$i=1,2,\ldots$, where we put $z_0=1$. The equations are invariant
under sign change of the $z_i$.
When $z_m=0$, the equations for $z_k$ with $k<m$ and $k>m$
decouple from each other. We have $z_{m-1}=1$ and $z_{m+1}=1$,
both up to a sign. But let us concentrate on the admissible case where no
$z_i$ vanishes. Then a priori the $z_i$ become rational functions
of $z_1$.

The reader who does not know this recursion will find it
somewhat miraculous that the $z_i$ turn out to be polynomials
in $z_1$ with integer ocefficients. Indeed
$z_i = p_i(z)$, where the $p_i$ are the standard Chebysheff polynomials
and we have put $z_1=z$. These polynomials are defined by a linear recursion,
namely
\beq{Cheblin}
p_{i+1}(z)+p_{i-1}(z) = zp_i(z)
\eeq
with $p_0(z)=1$ and $p_1(z)=z$. By 
they satisfy the quadratic recursion required for the $z_i$. Indeed
\begin{eqnarray*}
1 + p_{i-1}p_{i+1} = 1 + (zp_i-p_{i-1}) p_{i-1}  &=&\\
1 - p_{i-1}^2 + p_{i-2}p_i +  (zp_{i-1} - p_{i-2})p_i &=&
p_i^2.
\end{eqnarray*}
For $z=\omega + \omega^{-1}$ one easily obtains by induction from \req{Cheblin} 
$$p_i(z) = \omega^i + \omega^{i-2} + \ldots + \omega^{-i}$$
thus
$$p_i(z) = \frac{\omega^{i+1}-\omega^{-i-1}}{\omega -\omega^{-1}}$$
for $\omega\neq 1,-1$. 

For later use we will define polynomials $p_i(z)$ for negative $i$
by extending the linear recursion relation to such negative values.
This is easy, since the recursion relation is invariant under a sign
change of $i$. We obtain $p_{-1}(z)=0$, $p_{-2}(z)=-1$ and more generally
$p_{-i}(z)=-p_{i-2}(z)$. Note that 2 is the Coxeter number of $A_1$.

The solution of $U=AV$ for $A(A_1,A_n)$ can be obtained by
specializing the previous result to $p_{n+1}(z)=1$. This yields
$p_n(z) p_{n+2}(z) = 0$. The case $p_n(z)=0$ is not admissible. Thus we
obtain
$$\omega^{2(n+3)} = 1,$$
excluding $\omega^2=1$. This is the expected result, since
$$h(A_1)+h(A_n) = n+3.$$ 
Let us denote the resulting value of $p_i(z)$ by $z_i$. Since $z_{n+1}=1$
and $z_{n+2}=0$, the linear recursion yields $z_{n+3}=-1$ and more
generally $z_{i+n+3}=-z_i$. Thus we have an anti-periodic behaviour
with period $n+3$, which is the sum of the Coxeter numbers of $A_1$
and $A_n$. This behaviour generalizes to all ADET Lie algebras. 
Together with $p_{-i}(z)=-p_{i-2}(z)$ it yields $z_i=z_{n+1-i}$.

Now consider $A(A_1,D_{n+2})$. We write $Z=(z_1,z_2,\ldots, z_n,z',z'')$,
where $z',z''$ are the variables for the vertices on the short arms of
the Dynkin diagram of $D_{n+2}$. One obtains the equations
\begin{eqnarray*}
1 + z_{i-1}z_{i+1} &=& z_i^2\quad\mbox{for}\quad i=1,2,\ldots,n-1\\
1 + z_{n-1}z'z''  &=& z_n^2\\
z'^2=z''^2 &=& 1+z_n
\end{eqnarray*}
This yields $z_k=p_k(z)$, $z'z''=p_{n+1}(z)$ and $z''=\pm z'$. The equations
$1+p_n(z) = \pm p_{n+1}(z)$ are of degree $n+1$ and all solutions are found
easily when we put $z=\omega + \omega^{-1}$ as before.
Solutions are obtained from $\omega = \pm 1$ or $\omega^{n+1} = \pm 1$ or
$\omega^{n+2} =1$. The latter two possibilities are not admissible, however,
since they lead to $z_n=0$ or to $z'=z''=0$. For $\omega = \pm 1$ the choice 
of sign is irrelevant. One obtains
$$\exp(u_k) = (k+1)^{-2}$$ 
for $k=1,\ldots,n$ and $\exp(u')=\exp(u'')=(n+2)^{-1}$.

The corresponding central charge is $c_{eff}=1$, independently
of $n$. This follows easily by induction in $n$ and insertion of $2u=u_k$ 
in the doubling formula
$$2L(u,v) = 2L(u-v',v') + L(2u,v+v')$$
valid for $\exp(v')=\exp(u)+1$. The doubling formula is obtained from eq. 
\req{5t} by the identification $v_2=v_4=u$ and use of $L(u,v)+L(v,u)=\pi^2/6$.

The case $A(A_1,E_{n+3})$ is a bit more complicated.
We use variables $z_1,\ldots,z_n$ for the longest branch, plus
$u_1,u_2,t$ for the short branches. We put $u=u_1$, $z=z_1$.
The variable $t$ is determined by $t^2 = 1 + z_n$.
When none of the variables vanishes, we have $z_k=p_k(z)$,
but also $p_3(u)=u^3-2u=p_n(z)$ and
$$tu_2 = t(u^2-1) = p_{n+1}(z).$$
Eliminating $t,u$ is now easy. The resulting polynomial equation for $z$
has many unacceptable solutions for which some $p_k(z)$,
$k=1,\ldots,n+1$ vanishes.
The remaining solutions are given for $n=3$ by
$$z^3-2z^2-z+1 = 0$$
thus $z=\omega+\omega^{-1}+1$ with $\omega^7=1$,
for $n=4$ by
$$(z^2 - 3)^2 = 5$$
thus $z=\sqrt{2}(\sqrt{5}+1)/2$ up to sign choices,
and for $n=5$ by
$$(z-1)^2 = 2.$$

As next case let us consider $A=A(T_1,A_\infty)$, with
$A_\infty$ as before. This time we obtain the recursion relation
$$z_i + z_{i-1}z_{i+1} = z_i^2.$$
We put $z_k=q_k(z)$. The miracle repeats, with new polynomials. They satisfy 
the recursion relation
$$q_{i+1}(z) = zq_i(z)-q_i(z)-q_{i-1}(z)+1.$$
With $z=\omega + 1 +\omega^{-1}$ we find
$$z_i = \frac{(\omega^{(i+2)/2}-\omega^{-(i+2)/2})
(\omega^{(i+1)/2}- \omega^{-(i+1)/2})}{(\omega-\omega^{-1})
(\omega^{1/2}-\omega^{-1/2})},$$
where a limit has to be taken for $\omega^2=1$.
For $A=A(T_1,A_n)$ we have to impose $z_{n+1}=1$, $z_{n+2}=0$
as before. This yields
$$\omega^{n+4} = 1.$$
Moreover, only primitive $(n+4)$-th roots of unity are
admissible, since otherwise $z_i=0$ for some $i\leq n$.

Now let us consider $A=A(A_2,A_\infty)$. We write $z_{1i}=x_i$,
$z_{2i}=y_i$ and obtain the equations
\begin{eqnarray*}
x_i^2 &=& y_i + x_{i-1}x_{i+1}\\
y_i^2 &=& x_i + y_{i-1}y_{i+1} 
\end{eqnarray*}
for $i=0,1,\ldots$, with $x_0=y_0=1$ and $x_i=y_i=0$ for $i<0$. 
The miracle repeats and we find
\begin{eqnarray*}
x_{i+1} &=& xx_i - yx_{i-1} + x_{i-2}\\
y_{i+1} &=& yy_i - xy_{i-1} + y_{i-2}
\end{eqnarray*}
for $i=0,1,\dots$, with $x=x_1$, $y=y_1$.

For $x_i=y_i$ the equations reduce to the case $A(T_1,A_\infty)$.
To prove the result in general we use the additive recursion as
definition of the $x_i,y_i$ and prove the quadratic recursion
formula. The latter is true for $i=-1,0,1$. For $i\geq 1$ we have

\begin{eqnarray*}
x_{i+1}^2-x_ix_{i+2} &=&  x_{i+1}(xx_i-yx_{i-1}+ x_{i-2})
- x_i(xx_{i+1}-yx_i+x_{i-1})\\
&=& y(x_i^2-x_{i-1}x_{i+1}) -x_ix_{i-1} + x_{i+1}x_{i-2}\\
&=& yy_i -x(x_{i-1}^2-x_ix_{i-2}) + x_{i-2}^2 -x_{i-3}x_{i-1}\\
&=& yy_i -xy_{i-1} + y_{i-2} = y_{i+1}.
\end{eqnarray*}
Interchange of $x,y$, i.e. the symmetry of $A_2$ yields the 
other quadratic recursion formula.

We can use the linear recursion relation to define $x_i,y_i$
for negative values of $i$. We have

$$x_{i-2} = yx_{i-1} - xx_i + x_{i+1}$$
This yields $x_i=y_i=0$ for $i=-1,-2$ and $x_{-i}=y_{i-3}$.
When $x_{n+1}=y_{n+1}=1$ and $x_{n+2}=y_{n+2}=0$ as required
for $U=AV$ and $A=A(A_2,A_n)$ we obtain in addition
$x_{n+4+i}=x_i$, $y_{n+4+i}=y_i$. Again, $n+4$ is the sum of the
Coxeter number of $A_2$ and $A_n$. The result also applies to
the special case $A(T_1,A_n)$, since $h(T_1)=h(A_2)$.

So far, we always obtained real values for the solutions of our
algebraic equations. Now let us apply the previous equations
to the case $A(A_2,A_3)$.
We write all $x_i,y_i$ as polynomials in $x,y$ and have to
impose the algebraic conditions $x_4=y_4=1$. This yields
$x_3x_5=y_3y_5=0$, thus $x_5=y_5=0$ by admissability. Let us first
remark that explicit solvability in terms of roots of unity
can only be expected in the admissible case. 
When $y=0$, but no other component vanishes, we obtain
$x^5+x^3+2x^2-x+1=0$, an equation which does not seem to have
magical properties. 

We can assume
that $x\neq y$ since otherwise we are in the $A(T_1,A_3)$ case
which has been treated above. The polynomials $(x_4-y_4)/(x-y)$,
$x_4+y_4-2$, $(x_5-y_5)/(x-y)$ are symmetric under interchange
of $x,y$ and can be expressed in the variables $xy$ and $z=x+y$.
This yields $z^3-z+2-xy(2z+3)=0$, $(xy)^2-xy(z^2+1)+z^2-1=0$,
$(xy)^2-xy(3z^2+4z+3)+z^4+3z+2=0$. By elimination of $xy$ it is
easy to see that the only common solution is $z=-1$, $xy=2$.
This yields $(2x,2y)=(-1+i\sqrt{7},-1-i\sqrt{7})$, up to
complex conjugation.

Logarithms can be taken such that the relation $U=AV$ is
satisfied. For example one may take
\begin{eqnarray*}
\log(x_1) &=& \frac{\log(2)-(\pi +\alpha)i}{2}\\ 
\log(x_2) &=& -\pi i\\
\log(x_3) &=& \frac{\log(2)+(\pi +\alpha)i}{2} - 2\pi i 
\end{eqnarray*}
and $log(y_i)=log(x_{4-i})$.
Applying Roger's dilogarithm to the resulting
torsion element $\sum_i(u_i,v_i)$ of the Bloch group one finds

$$2\,\Re\, L\left(\frac{3+i\sqrt{7}}{8}\right) = 
\pi\left(\arg\left(\frac{3+i\sqrt{7}}{8}\right)+
         \arg\left(\frac{5+i\sqrt{7}}{8}\right)\right)
+\pi^2/4.$$
The phases arise, because it is not possible to choose $\log(x)$
complex conjugate to $\log(y)$.

As last example before the the introduction of some Lie algebra theory 
let us consider the case $A=A(A_3,A_\infty)$. For aesthetic
reasons we label the variables as $z_{1i}=x_i$, $z_{2i}=y_i$, $z_{3i}=z_i$ 
without worrying about the previous use of $z_i$. The equations are
\begin{eqnarray*}
x_i^2 &= y_i + x_{i-1}x_{i+1}\\
y_i^2 &= x_iz_i + y_{i-1}y_{i+1}\\
z_i^2 &= y_i + z_{i-1}z_{i+1}
\end{eqnarray*}
with $x_0=y_0=z_0=1$. We put $x=x_1$, $y=y_1$, $z=z_1$.
We will see that in the admissible case
$$x_{i+1}= xx_i - yx_{i-1} + zx_{i-2} -x_{i-3}$$
for $i=1,2,\ldots$, with $x_i=0$ for $i<0$. Of course $x,z$
can be interchanged by the symmetry of $A_3$. The recursion for
$y_i$ is more complicated. One obtains
\begin{eqnarray*}
y_{i+1} &=& yy_i +xzy_{i-1} - yy_{i-2} + y_{i-3}\\
         && + (x^2+z^2)(y_{i-2}+y_{i-4}+y_{i-6}+\ldots)\\
         && - 2xz(y_{i-1}+y_{i-3}+y_{i-5}+\ldots)
\end{eqnarray*}

We show by induction that these additive recursion formulas imply the
multiplicative ones, assuming the latter for $j\leq i$.
The proof is somewhat tedious to follow. As before, one has to substitute 
the additive recursion formulas for the $x_i$ at many places of the
equations. To indicate where the substitutions happen we write
$\xi=x$ at the relevant positions. One has
\begin{eqnarray*}
x_{i+1}^2-x_ix_{i+2} &=& x_{i+1}(xx_i -yx_{i-1} +zx_{i-2} -x_{i-3})\\
                      &&\ \ \  -x_i(xx_{i+1}-yx_i + zx_{i-1} - x_{i-2})\\  
&=& y(x_i^2-x_{i-1}x_{i+1}) +z(\xi_{i+1}x_{i-2}-\xi_ix_{i-1})\\
   &&\ \ \  -\xi_{i+1}x_{i-3} + \xi_ix_{i-2}\\
&=& yy_i - xzy_{i-1} + z^2y_{i-2}+z(x_{i-4}x_{i-1}-x_{i-3}x_{i-2})\\
  &&\ \ \ -x(x_ix_{i-3}-x_{i-1}x_{i-2})-yy_{i-2}+y_{i-3}.
\end{eqnarray*}

To prove that the last expression is equal to $y_{i+1}$ one has to show
\begin{eqnarray*}
&z&(x_{i-1}x_{i-4}-x_{i-2}x_{i-3})
  - x(x_ix_{i-3}-x_{i-1}x_{i-2})= \\
&&x^2(y_{i-2}+y_{i-4}+y_{i-6}+\ldots) - 2xz(y_{i-3}+y_{i-5}+\ldots)
+z^2(y_{i-4}+y_{i-6}+\ldots)
\end{eqnarray*}
The brackets on the left hand side have identical structures,
except for a shift of the indices by 1. Thus it is sufficient to
calculate one of them. One finds
\begin{eqnarray*}
\xi_ix_{i-3}-\xi_{i-1}x_{i-2} 
&=& -xy_{i-2}+zy_{i-3}+ (\xi_{i-2}x_{i-5}-\xi_{i-3}x_{i-4})\\
&=& -x(y_{i-2}+y_{i-4}+\ldots) + z(y_{i-3}+y_{i-5}+\ldots)
\end{eqnarray*}
where an obvious induction has been used in the last step. This
finishes the proof of the formula for $x_{i+1}^2$. Interchange of $x,z$
yields the one for $z_{i+1}^2$. For $y_{i+1}^2$ a somewhat different
approach is easier.  We use the equality $x_i^2 = y_i + x_{i-1}x_{i+1}$
to express $y_i$ in terms of the $x_j$. This yields

$$y_i^2-y_{i-1}y_{i+1} = x_iP_i$$
where 
$$P_i = x_i^3-2 x_{i-1}x_i x_{i+1}+ x_{i-1}^2 x_{i+2}
   + x_{i-2} x_{i+1}^2 -  x_{i-2}x_ix_{i+2}.$$
Thus it suffices to show $P_i=z_i$ by induction. Now
\begin{eqnarray*}
P_{i+1} &=&
\xi_{i+3}(x_i^2- x_{i-1} x_{i+1})+\xi_{i+2}( x_{i-1} x_{i+2}-x_i x_{i+1})\\
&&+\xi_{i+1}( x_{i+1}^2-x_i x_{i+2})\\
&=& zP_i + \xi_{i+2}(x_ix_{i-3}-x_{i-1}x_{i-2})
  + \xi_{i+1}(x_{i-1})^2-x_{i+1}x_{i-3})\\
 && + \xi_i(x_{i-2}x_{i+1}-x_ix_{i+2})\\
&=& zz_i -yP_{i-1} + \xi_{i+1}(x_{i-3}^2-x_{i-2}x_{i-4})
  + \xi_i(x_{i-1}x_{i-4}-x_{i-2}x_{i-3})\\
 && + \xi_{i-1}(x_{i-2}^2-x_{i-1}x_{i-3})\\
&=& zz_i-yz_{i-1} + xz_{i-2} - z_{i-3} = z_{i+1}
\end{eqnarray*}
which finishes the proof.
 
Applying the linear recursion relation to negative values we obtain
$x_{-i}=-z_{i-4}$ and $y_{-i}=-y_{i-4}$. For solutions of $U=AV$
with $A=A(A_3,A_n)$ we also find the periodicity $x_{i+n+5}=x_i$
and analogously for $y,z$.

At this stage the reader should be convinced that elementary algebra
is not entirely satisfactory for an understanding of what is going on.
Obviously, the generalisation of the linear recursion for $x_i$ to
$A(A_m,A_\infty)$ is easy to guess,. and maybe even to prove. When one
looks at the coefficients, one gets a path through the Dynkin diagram of $A_m$
which starts at  one side and ends at the other. This is one of Ocneanu's
essential paths \cite{O99}, namely the one which starts at the $x$-vertes. For  
the $y_i$ terms in $A(A_3,A_\infty)$ we first go from $y$ to $x,z$ and then 
back to $y$. Again this agrees with Ocneanu's essential path starting
at the $y$-vertex. Maybe this generalizes, but it doesn't give us the
chain of correction terms, which certainly will become more complicated
for Lie algebras of higher rank.

Thus we will leave this approach and jump to conclusions which for us come a bit 
out of the blue, but do work \cite{K87, KR87}. We need another interpretation of 
the algebraic equations which gives a new meaning to the solutions. Since
the latter are sums of roots of unity with integral coefficients, one can 
expect that these integers count something. For the Chebysheff polynomials 
we have $p_i(2)=i+1$, numbers which will be interpreted as the dimensions
of the irreducible representations of $A_1$. Dimensions can be obtained
as character values at the unit element, and more generally the algebraic
integers will be given as character values at elements of finite order.
Apart from some general remarks, we shall only consider the case $A(A_m,A_n)$,
where it suffices to work with Lie groups instead of quantum groups.
In this case the argumentation will be quite self-contained, apart from the 
fact that the reader is supposed to know or look up the Littlewood-Richardson 
rules. On the other hand, we briefly recall most of the relevant facts of
representation theory. The Weyl character formula is mentioned to provide
some context, but its relevant special cases are discussed without reference
to the general theory. We will work over the complex numbers and in the 
context of algebraic groups, which avoids many unnecessary complications.

Let us first consider the irreducible finite dimensional representations
of $GL(n)$. In particular, the subgroup of diagonal matrices, with
elements $g=diag(a_1,\ldots,a_n)$ will be represented. Its irreducible
representations are one-dimensional and map $g$ to $\prod_{k=1}^n a_k^{i_k}$, 
with integral $i_k$. Such a representation is called a weight and
notated as a sequence $I=(i_1,\ldots,i_n)$. One-dimensional representations
can be multiplied and inverted. In terms of the sequences $I$ this
group structure is additive - the weights form a lattice.
When restricted to the diagonal elements, the representations $\rho$ of 
$GL(n)$ decompose into a finite number of weights. The diagonal matrices
$g=diag(a,\ldots,a)$ form the center of $GL(n)$. They are represented
by $a^{|I|}$, where $|I|=i_1+\ldots+i_n$. When $\rho$ is irreducible,
the center is represented by multiples of the identity matrix, such that
$|I|$ takes the same value for all weights occuring in $\rho$. 

Let the $GL(n)$ representations $\rho$, $\rho'$ of dimensions $d,d'$ decompose
into weights $I_r$, $r=1,\ldots,d$ and  $I'_s$, $s=1,\ldots,d'$.
Then the tensor product $\rho\otimes\rho'$ decomposes into weights
$I_r+I'_s$. The symmetric power $S^k\rho$ decomposes
into weights $I_{r_1}+\ldots+I_{r_k}$ with $r_1\geq\ldots\geq r_k$
and the external power $\Lambda^k\rho$ decomposes into weights
$I_{r_1}+\ldots+I_{r_k}$ with $r_1 >\ldots >r_k$. 

The character $\chi:\ GL(n)\rightarrow\C$ of a representation $\rho$
of dimension $d$ is given by $\chi(g)=Tr\rho(g)$. It has the form
$$\chi(g)=\sum_I\prod_{k=1}^n a_k^{i_k}, $$
where the sum goes over the $d$ weights occuring in $\rho$. With
a sequence $I=(i_1,\ldots,i_n)$ all its permutations appear, too, since
permutations of the entries of $g$ can be achieved by conjugation in $GL(n)$.
When one gives the standard lexicographic order to the $(i_1,\ldots,i_n)$,
any irreducible representation of $GL(n)$ has a unique highest weight.
Since any sequences $(i_1,\ldots,i_n)$ can be permuted to a decreasing one,
such a highest weight is given by a decreasing sequence of $n$ integers. One
can show that all such sequences can occur as highest weights. This yields
a complete classification of the isomorphism classes of irreducible $GL(n)$
representations. We write the corresponding representation as $\rho_I$,
its character as $\chi_I$.

The character of the defining representation of $GL(n)$ maps $g$ to
$Tr(g)=\sum_{i=1}^n a_n$. Thus its weights are $(1,0,\ldots,0)$ and
its permutations, and its highest weight is $(1,0,\ldots,0)$.
The $k$-th symmetric product of this representation is irreducible and has 
highest weight $(k,0,\ldots,0)$. Its  $k$-th exterior product is irreducible
and has highest weight $(\underbrace{1,\ldots,1}_{k\ terms},0,\ldots,0)$. 
For $k>n$ the exterior product is 0, for $k=n$ it yields the one-dimensional
determinant representation, with unique weight $(1,\ldots,1)$.
For $I=(\underbrace{i,\ldots,i}_{k\ terms},0,\ldots,0)$ we write
$\rho_I=\rho_i^k$ and $\chi_I=\chi_i^k$.

The character of the irreducible $GL(n)$ representation with 
highest weight $I=(i_1,\ldots,i_n)$ is given by
$$\chi_I(g)= D_I(g)/D_0(g), $$
where $D_I(g)=\det(M^I)$ and the matrix $M^I$ has entries
$$M^I_{lr}=a_r^{i_l+n-l},$$
$l,r=1,\ldots,n$. This is a special case of the Weyl character formula.
For $I=(\underbrace{i,\ldots,i}_{k\ terms},0,\ldots,0)$ we often write
$\rho_I=\rho_i^k$ and analogously for $\chi_I$ and $M^I$. 

Since $\rho_i^1$ is the $i$-th symmetric product of $\rho_1^1$, one has 
$$\chi_i^1(g)=
\sum_{m_1,\ldots,m_n\in\N \atop m_1+\ldots m_n=i}\prod_{r=1}^n a_r^{m_r}. $$
To check the Weyl character formula in this case, one first notes that 
$D_0$ is a Vandermonde determinant, such that
$$D_0(g)=\prod_{r<s}(a_r-a_s).$$
In our case the matrix $M^I$ agrees with $M^0$ apart from the first row,
which has entries $a_r^{i+n-1}$. Its determinant can de developed in
terms of minors with respect to the first row. Up to a sign, the minor 
multiplying $a_r^{i+n-1}$ is $D_0(g_r)$, where the $g_r$ are diagonal
matrices in $GL(n-1)$ and arise from $g$ by supressing $a_r$. One can
that this development agrees with $D_0(g)\chi_i^1(g)$. Indeed
$$(a_1-a_2)\chi_i^1(g)=\chi_{i+1}^1(g_2)-\chi_{i+1}^1(g_1).$$
By induction one sees immediately that
$$\prod_{r=2}^s(a_1-a_r)\chi_i^1(g)=\chi_{i+s-1}^1(g^s)+R_s,$$
where the $g^s$ are diagonal matrices in $GL(n-s+1)$ which arise from $g$ by 
supressing $a_2,\ldots,a_s$ and the $a_1$ degree of $R_s$ is at
most $s-2$. Using this result for $s=n$, where $\chi^{i+n-1}(a_1)=a_1^{i+n-1}$, 
one sees that $D_0(g)\chi_i^1(g)-D_i^1(g)$ has degree at most $n-2$ in $a_1$
and by permutation symmetry in all $a_r$. When one looks at the developments
in first row minors of $D_i^1(g)$ and $D_0(g)$, one sees that such terms
cannot occur, such that $D_0(g)\chi_i^1(g)-D_i^1(g)=0$. 
 
When one restricts $GL(n)$ to $SL(n)$, the determinant representation becomes
trivial. Thus weights $(i_1,\ldots,i_n)$ and $(i_1+j,\ldots,i_n+j)$, $j\in\Z$,
become equivalent, and one can regard the sequences $\lambda= 
(i_1-i_2,\ldots,i_{n-1}-i_n)$ as the weights of $SL(n)$. All such integral
can occur, since each irreducible representation of $SL(n)$ can be lifted to 
$GL(n)$. For the highest weights of $SL(n)$ representations, all entries are
non-negative. These highest weights form a semi-group which is generated
by the $n-1$ fundamental weights $(1,0,\ldots,0)$, $(0,1,0,\ldots,0)$, 
$\ldots$ $(,0,\ldots,1)$. The corresponding representations are just the
exterior products of the defining representation mentioned above, with
$k=1,\ldots,n-1$. These are the fundamental representations of $SL(n)$.
The Dynkin diagram of $SL(n)$ is an ordered chain of $n-1$ vertices, which
can be labelled by $k=1,\ldots,n-1$. Thus each of them corresponds to one
of the fundamental representations, and each irreducible representation of 
$SL(n)$ can be classified up to isomorphism by associating natural numbers
$i_k-i_{k+1}$ to the corresponding vertices.

The Dynkin diagram $A_{n-1}$ of $SL(n)$ has an obvious reflection
symmetry. In terms of the characters this yields in particular
$$\chi_i^k(g^{-1})=\chi_i^{n-k}(g).$$

Instead of the Lie groups $GL(n)$ and $SL(n)$ one can consider their
Lie algebras. In both cases, the Lie algebra of the subgroup of diagonal 
matrices forms a maximal abelian subalgebra. The classification of the
irreducible representations of the Lie algebras is the same as for the
Lie groups. 

The procedure generalizes to the other simple Lie algebras $X$ and their finite
dimensional representations.  As for $GL(n)$ and $SL(n)$ 
one chooses a Cartan subalgebra, i.e. a maximal abelian subalgebra of $X$.
Its irreducible representations are called weights and form a lattice $\Lambda$
isomorphic to $\Z^r$, where $r$ is the rank of $X$. The vertices of the
Dynkin diagram of $X$ correspond to a basis of the dual lattice of $\Lambda$,
such that weights yield an integer for each vertex. The Cartan matrix
corresponding to the Dynkin diagram yields a metric on the dual of $\Lambda$
and therefore on $\Lambda$ itself. 

The weight lattice
can be ordered in some lexicographic way. The irreducible
representations of $X$ have highest weights, which give a natural number
(possibly zero) for each vertex of the Dynkin diagram. This yields
a classification of the isomorphism classes of these representations.
In particular, they form a semi-group. Choosing zero for each vertex 
yields the trivial one-dimensional representation, the zero of this semigroup.
Choosing one for some vertex $k$ and zero for the others yields the fundamental
representations with highest weights $\lambda^k$. 

The tensor product $\rho\otimes\rho'$ of two irreducible representations
$\rho$, $\rho'$
decomposes into a direct sum of irreducible representations. Among their
weights one is the highest, and given by the sum of the highest weights
of $\rho$ and $\rho'$.
For a representation $\rho$ the corresponding character $\chi:X \rightarrow \C$
is given by $\chi(x)= Tr \rho(x)$. We also use $\chi$ for the Lie group
representations obtained by exponentiating the $\rho(x)$. The character of
$\rho\otimes\rho'$ is the ordinary product $\chi\chi'$.
The dimension of $\rho$ is equal to $\chi(1)$. 

After this extensive recall of Lie algebra theory we can come back to our 
algebraic equations. For $X=A_1$ the Dynkin diagram has one
vertex only, so the irreducible representations $\rho_i$ are classified by
a single natural number $i=0,1,2,\ldots$. The dimension of $\rho_i$ turns out
to be $i+1$. We have $\chi_i=p_i(\chi_1)$, where the $p_i$ are the Chebysheff 
polynomials. A check can be made by evaluating this relation at $g=1$,
where one indeed has $i+1=p_i(2)$, which is the correct dimension. To prove
the formula one can use the well-known tensor product relation
$$\rho_1\otimes\rho_i = \rho_{i-1}\oplus \rho_{i+1}$$     
for $i>0$, which yields
$$\chi_1\chi_i=\chi_{i-1}+\chi_{i+1}. $$
This agrees with the recursion relation for the $p_i(x)$. Since also
$p_0(\chi_1)=\chi_0=1$ and $p_1(\chi_1)=\chi_1$, it is clear that
$\chi_i=p_i(\chi_1)$. The multiplicative recursion relation for the $p_i$
can be explained in the same way, since
$\rho_i\otimes\rho_i = \rho_0\oplus\rho_2\oplus\cdots\oplus\rho_{2i}$
and $\rho_{i-1}\otimes\rho_{i+1}=\rho_2\oplus\cdots\oplus\rho_{2i}$.

For $X=A_m$, let us consider the $GL(m+1)$ representations $\rho_i^k$
introduced above. Like all irreducible representation of $GL(m+1)$, they
stay irreducible when the representation is restricted to $SL(m+1)$ or its
Lie algebra $A_m$. Indeed, the elements of $GL(m+1)$ are products of
$SL(m+1)$ elements and a elements in the center, and in irreducible
representations the latter  are represented by multiples of the identity 
operator. The $\chi_i^k$ can be written as polynomials in the $\chi_1^k$.
It turns out that these polynomials coincide with the ones we obtained for
$A(A_m,A_\infty)$. Moreover we can write $z_{ki}=\chi_i^k(g)$, where 
$g\in SL(m+1)$ is determined up to conjugation by $z_{k1}=\chi_1^k(g)$.

To prove this result we must to show that for $i=1,\ldots,m$
and $k=1,2,\ldots$ one has
$$\rho_i^k\otimes \rho_i^k = \left(\rho_{i+1}^k\otimes\rho_{i-1}^k\right)
\oplus \left(\rho_i^{k+1}\otimes\rho_i^{k-1}\right).$$
Here the $\rho_0^k$ and the $\rho_i^0$ are triviall one-dimensional 
representations. The $\rho_i^{m+1}$ are the $i$-th powers of the
determinant representation, which becomes trivial for $g\in SL(m+1)$.

The formula follows immediately from the Littlewood-Richardson rules for
tensor products of irreducible representations of $GL(m+1)$, but here only 
a brief sketch of the derivation will be given. As we have seen, these 
representations are classified by decreasing sequences $I=(j_1,\ldots,j_{m+1})$ 
of integers. The one-dimensional determinant representation is given by
$(1,\ldots,1)$ and its tensor product with the representation of highest
weight $(j_1,\ldots,j_{m+1})$ yields the highest weith
$(j_1+1,\ldots,j_{m+1}+1)$. 

We are only interested in representations with $j_{m+1}\geq 0$. For
notational convenience we represent the corresponding highest weights as 
infinite sequences $I=(j_1,\ldots,j_{m+1},0,0,\ldots)$. We define the length 
$l(I)$ of $I$ as the number of its non-zero terms and we let all sequences
with length $l(I)>m+1$ correspond to the 0 representation. Then the $\rho_i^k$
$(\underbrace{i,\ldots,i}_{k\ \ times},0,\ldots)$.
The Littlewood-Richardson rules imply that
$\rho_i^k\otimes \rho_i^k$ is the direct sum of the representations
given by  $(i+j_1,\ldots,i+j_k, i-j_k, \ldots, i-j_1,0,\ldots)$,
where $(i,j_1,\ldots,j_k,0)$ is a decreasing integral sequence. Similarly
$\rho_{i+1}^k\otimes\rho_{i-1}^i$ corresponds to the direct sum of
the subset of these representations for which $j_k=0$ and
$(\rho_i^{k+1}\otimes\rho_i^{k-1}$ corresponds to the complementary subset. 
This proves our equality. 

The linear recursion relations for the $\rho_i^1$ are easy to prove, too.
According to the Littlewood-Richardson relations the tensor product
of $\rho_i^1$ with $\rho_1^k$ is given by the sum of two irreducible
representations with highest weights
$(i+1,\underbrace{1,\ldots,1}_{k-1\ \ times})$ and
$(i,\underbrace{1,\ldots,1}_{k\ \ times})$. This immediately yields
\begin{eqnarray*}
\left(\rho_i^1\otimes \rho_1^1\right)\oplus 
\left(\rho_{i-2}^1\otimes \rho_1^3\right) 
\oplus \left(\rho_{i-4}^1\otimes \rho_1^5\right)\oplus\dots=\\
\rho_{i+1}^1\oplus \left(\rho_{i-1}^1\otimes \rho_1^2\right)
\oplus \left(\rho_{i-3}^1\otimes \rho_1^4\right)\oplus\dots 
\end{eqnarray*}
which in turn yields for $i\geq m$. 

$$\chi_{i+1}^1=\chi_1^1\chi_i^1-\chi_1^2\chi_{i-1}^1+\ldots
+(-)^m\chi_1^{m+1}\chi_{i-m}^1. $$
This formula can be interpreted as a recursion relation for the $\chi_i^1$.
It stays true for $i=0,\ldots,m-1$, if one puts $\rho_i^1=0$ for
$i=-1,-2,\ldots,-m$. This result agrees with what one obtains from a 
continuation of the Weyl character formula for $\chi_i^1$ to negative 
values of $i$. Indeed the matrices $M_i^1$ have two equal rows for
$i=-1,-2,\ldots,-m$, such that their determinants vanish. 

We have seen that for $g\in SL(m+1)$ the multiplicative recursion relations
for the $\chi_i^k(g)$ agree with those of the $z_{ki}$. For admissible
solutions of the algebraic equations, the $z_{ki}$ are uniquely
determined by the $z_{k1}$. Thus it remains to show that every choice of
the $z_{k1}$ can be parametrised in the form $z_{k1}=\chi_1^k(g)$.
Since the $\chi_1^k(g)$ are just the elementary symmetric polynomials of
the $a_1,\ldots,a_{m+1}$ this is indeed possible. When one leaves
$z_{m+1,1}$ free one has a bijection between the $z_{k1}$ and the 
$a_1,\ldots,a_{m+1}$, the latter taken over $\C$ and up to permutations.
The condition $z_{m+1,1}=1$ restricts $\diag(a_1,\ldots,a_{m+1})$ to
$SL(m+1)$.

To find admissible solutions of our algebraic equations for
$A(A_m,A_n)$ we need $\chi_{n+1}^k(g)=1$, $\chi_{n+2}^k(g)=0$ for
$k=1,\ldots,m$. We first will show that this implies
$\chi_{n+1+l}^k(g)=0$ for $l=1,\ldots,m$.

More generally, for such $l$ we will show $\chi_I(g)=0$, if $j_1>n+1$, 
$j_1+l(I)= n+l+2$, $|I|\geq l+l(I)(n+1)$. Here $\chi_I$ is the character
of the representation with highest weight $I=(j_1,j_2,\ldots)$.
Recall that $|I|$ is the sum of the $j_k$. Note that the inequalities
imply $l(I)\leq m$.

The proof works by induction in $l,j_1$, in lexicographic order. For
the smallest allowed value $j_1=n+2$ we have $l(I)=l$. This yields
$|I|\leq l(n+2)$, since $I$ is a decreasing sequence. On the other hand
$|I|\geq l+l(I)(n+1)= l(I)(n+2)$. Thus the inequalities are saturated
and 
$$I=(\underbrace{n+2,\ldots,n+2}_{l\ terms},0,\ldots).$$
Thus $\chi_I=\chi_{n+2}^l$ and we have nothing to prove.

In general, let $I$ be strictly decreasing after exactly $r$ terms,
such that $j_r=j_1$, $j_{r+1}<j_1$. For $r=l$ we have $l(I)=l$ and
$j_1=n+2$, so we are in the previous case. Thus we may assume $r<l$.
By assumption, this implies $j_1>n+2$.

Now the Littlewood-Richardson rules yield
\begin{eqnarray*}
\rho_I\otimes&& \rho_{n+1}^1\\ &&= (\rho_{I-\alpha_1}\otimes \rho_{n+2}^1)
\oplus - (\rho_{I-\alpha_2}\otimes \rho_{n+3}^1)\oplus\ldots
\oplus (-)^{r+1} (\rho_{I-\alpha_r}\otimes \rho_{n+1+r}^1)\\
&&\bigoplus_{I'} (\pm)\rho_{I'}
\end{eqnarray*}
where minus signs are interpreted in the form $\rho-\rho=0$, the $\alpha_k$
have the form
$$\alpha_k=(\underbrace{0,\ldots,0}_{r-k\ terms},
\underbrace{1,\ldots,1}_{k\ terms},0,\ldots)$$
and the highest weights $I'$ satisfy $j'_1=j_1-1$, $l(I')\leq l(I)+1$,
$|I'|=|I|+n+1$. In particular $j'_1>n+1$, $j'_1+l(I')\leq n+l+2$,
$|I'|\geq l+l(I')(n+1)$, such that we can use the induction hypothesis
$\chi_{I'}(g)=0$. By the induction hypothesis we also have
$\chi_{n+k+1}^1(g)=0$ for $k=1,\ldots,r$. Together with $\chi_{n+1}^1(g)=1$
this yields $\chi_I(g)=0$, and the proof is finished.

Next we will show that the algebraic equations imply that $D_0(g)\neq 0$,
such that the eigenvalues $a_r$ of $g$ must all be different. 
We assume $n\geq m$, otherwise we exchange $A_m$ and $A_n$.
We have
$$a_{n+1}\chi_{n+2}^1(g)-\chi_{n+3}^1(g)=\chi_{n+3}^1(g_{n+1}),$$
where $(g_{n+1}\in GL(n)$ is given by $g=\diag(a_1,\ldots,a_n)$.
By iteration we find that $\chi_{n+m+1}^1((a_1,a_2))$
is a linear combination of $\chi_{n+1+l}(g)$, $l=1,\ldots,m$ and thus
has to vanish. On the other hand
$$\chi_{n+m+1}^1((a_1,a_2))=a_1^{n+m+1}+a_1^{n+m}a_2+\ldots+a_2^{n+m+1}$$
and the right hand side cannot vanish for $a_1=a_2$, unless $a_1=0$, which
is not permitted by $g\in GL(n)$. By permutation
symmetry, all $a_r$ have to be different.

Use of the linear recursion relation for the $\chi_i^1$ allows to
express $\chi_{n+m+2}^1$ in terms of $\chi_{n+1+l}^1$ with $l=0,\ldots,m$.
This yields $\chi_{n+m+2}^1(g)=(-)^m$ for the solutions of our algebraic
equations, thus $D_{n+m+2}^1(g)=(-)^m D_0^1(g)$. More generally, 
$$D_{i+n+m+2}^1(g)=(-)^m D_i^1(g)$$
for $i=-m,\ldots,0$, since both sides vanish for $i=-m,\ldots,1$. 
Rows 2 to $m+1$ of the $M_i^1$ are given by the same Vandermonde
matrix. The first row has entries $a_k^{m+i}$. Developing the determinants
$D_i^1$ into minors with respect to the first row yields a system of $m+1$ 
linear equations for the $a_k^{m+n+2}$, $k=1,\ldots,m+1$. The determinant
of this system is a product of Vandermonde matrices, thus non-zero. This yields
the unique solution $a_k^{m+n+2}=1$ for all $k$.

These conditions are necessary for admissible solutions of our
algebraic equations. To show that they actually solve the equations we
just have to notice that they imply

$$D_{i+n+m+2}^k(g)=(-)^m D_i^k(g)$$
for all $k$. For $i=-m-1$ and $D_0(g)\neq 0$ this yields $\chi_{n+1}^k(g)=1$, 
such that the equations are satisfied. Admissibility has to be investigated 
separately, but this will not be done here.

We can formulate our result in the following way. Admissible solutions
of our algebraic eqations for $A(A_m,A_n)$ can be written as $\chi_i^k(g)$,
where $g$ is an element of $SU(m+1)$ which satisfies
$$g^{m+n+2}=(-)^m $$
and all of whose eigenvalues are different.
Note that $m+n+2$ must be the smallest power of $g$ which is equal to $1$
or $-1$, since for a small power $m+k+2$ with this property one finds
$\chi_i^{k+2}(g)=0$. The reader is encouraged to recover the solutions for
the special cases $A(A_1,A_m)$ and $A(A_2,A_3)$ discussed above from our
general result for $A(A_m,A_n)$.

Before we leave this case, we consider two symmetry properties of our
solutions. Applying the symmetry of the $A_m$ or $A_n$ diagram to any
solution yields another solution. We shall see that simultaneous application 
of these two operations leaves any solution invariant.

By the symmetry of the $A_m$ diagram, we have the same linear
recursion relation for the $\chi_i^m$, except that the coefficients
$\chi_1^k$ are replaced by $\chi_1^{m+1-k}$. For the solutions of our
algebraic equations this means that they obey linear recursion
relations which are invariant under simultaneous application of the
symmetries of $A_m$ and $A_n$. In the special cases $A=A(A_m,A_n)$, $m=1,2,3$
considered above we have found the same symmetry in all admissible solutions
themselves. We will prove that this remains true for all $m$.
The coefficients of the linear recursion relation for the $\chi_i^1(g)$
are the $\chi_1^k(g)$, $k=1,\ldots.m$. When one reads the relations in the
opposite sense, as relations among the $\chi_{-i}^1(g)$, the coefficients
become $\chi_1^{m+1-k}(g)=\chi_1^k(g^{-1})$. Use of the $A_m$ symmetry
yields linear recursion relations for the $\chi_{-i}^m(g)$, but now with
the original coefficients $\chi_1^k(g)$.

The linear recursion relations allow us to express $\chi_{i+m+1}^1(g)$
in terms of $\chi_{i+l}^1(g)$ with $l=0,\ldots,m$ and analogously 
$\chi_{i-m-1}^m(g)$ in terms of $\chi_{i-l}^m(g)$, with the same
coefficients. Now for admissible solutions of our algebraic equations we
have found $\chi_i^1(g)=\chi_{n+1-i}^m(g)$ for $i=-m,\ldots,0$. By
induction, the linear recursion relations yield the same equality for all
values of $i$, in particular the values $i=1,\ldots,m$ we need for the 
solutions of our algebraic equations. Moreover, the $\chi_i^1(g)$ or
$\chi_i^m(g)$ determined all $\chi_i^k(g)$, such that 
$$z_{ki}=z_{m+1-k,n+1-i}$$
for all $k,i$.

Finally one expects a duality between $A(A_n,A_m)$ and $A(A_m,A_n)$. Indeed
consider $g=diag(a_1,\ldots,a_{m+1})$ with $g^{m+n+2}=(-)^m$ and no
coinciding eigenvalues. The $a_i$ form $n+1$ of the $(m+n+2)$-th roots of
$(-)^m$. Let $(a_{m+2},\ldots,a_{m+n+2})$ be the remaining roots.
Then $h=diag(a_{m+2},\ldots,a_{m+n+2})$ satisfies $h\in SL(n+1)$ and
$h^{m+n+2}=(-)^m$, as one can check easily. In this way one can lift the
restriction $m\leq n$ we used to conclude that $g$ had no degenerate
eigenvalues. 

The calculations we made for $A(A_m,A_n)$ can be extended to $A(D_m,A_n)$
and $A(E_m,A_n)$, but one has to go beyond Lie algebras into the domain of
quantum groups, as shown in \cite{K87, KR87}.  Quantum groups have a co-algebra 
structure, which allows to define tensor products of representations as in
the Lie algebra case. For each simple Lie algebra
$X_m$ there is a quantum group $Y(X_m)$, called the Yangian of $X_m$,
which contains the enveloping algebra $U(X_m)$ of $X_m$. Recall that
the representations of $U(X_m)$ are given in a natural way by the
representations of $X_m$ itself.

For the special case $X_m=A_m$ there is a map from
$A_m$ to $Y(A_m)$ which reduces the representation theory of $Y(A_m)$
to the one of $A_m$, since every representation of $A_m$ can be
extended to a representation of $Y(A_m)$ on the same vector space,
and all irreducible representation of $Y(A_m)$ remain irreducible as
representations of $U(A_m)$. For $D_m$ and $E_m$ this no longer true.
Nevertheless the representations of $Y(X_m)$ can still be decomposed
into weights of $X_m$ and irreducible representations can be labelled
by their unique highest weight. Tensor products correspond to the
addition of weights as for $X_m$ itself. Let $\rho_i^k$ be the representation
of $Y(X_m)$ with highest weight $i\lambda^k$, where $k$ runs over the
vertices of the Dynkin diagram of $X_m$. As before, the $z_{ki}=\chi_k^i(g)$
can be written as polynomials of the $z_{k1}$ and satisfy the algebraic
equations of $A(X_m,A_\infty)$. Imposing $\chi_{n+1}^i(g)=1$,
$\chi_{n+2}^i(g)=0$ should allow to solve the algebraic equations for
$A(X_m,A_n)$. Indeed the special real solution relevant for calculating
the central charge $c$ is known and given by

$$g=\exp\left(-\frac{2\pi i\rho}{h(X_m)+n+1}\right), $$
where $\rho$ is the sum of all $\lambda^k$. Thus $g_0=g^{h(X)+n+1}$ lies
in the center of the simply connected compact Lie group corresponding
to $X$ and satisfies 
$$\rho_j^k(g_0)=\exp(-2\pi ij(\lambda_k,\rho),$$
where we have used the natural scalar product on $\Lambda$. One may
expect that all solutions of the algebraic equations arise in this way,
but the precise conditions on $g$ seem to be still unknown.

\section{Conclusions}
Of the unsolved problems mentioned in this article many are of a purely
mathematical nature and seem to be quite accessible. Much more challenging
is the development of a mathematically satisfactory theory of integrable
quantum field theories in two dimensions. The physics literature provides
a wealth or ideas and results, and perhaps it is time that mathematicians
get interested. Those theories which arise from perturbations of rational
conformal field theories may be the easiest ones to study.

\bibliography{leshouches}
 
\end{document}